\documentstyle[11pt]{article}
\newcommand \beq{\begin{eqnarray}}
\newcommand \eeq{\end{eqnarray}}
\newcommand \bib{\bibitem}
\newcommand \Tr{\mbox{\rm Tr}}
\begin{document}
\begin{quote}

\begin{center}

{\bf SCREENING EFFECTS ON THE ELASTIC NUCLEON-NUCLEON CROSS SECTION 
IN RELATIVISTIC NUCLEAR MATTER}

\vspace{1.5cm}

Joaquin DIAZ-ALONSO and Lysiane MORNAS \\
{\it D.A.R.C., Observatoire de Paris-Meudon, UPR 176 CNRS} \\
{\it F-92195 Meudon, France}

\vspace{2.cm}

{\bf ABSTRACT}

\end{center}

We investigate the screening effects on the nucleon-nucleon elastic cross
section inside nuclear matter at zero and finite temperature. The N-N
interaction is described phenomenologically via meson exchanges ($\sigma$,
$\pi$ and $\omega$) in the framework of a relativistic lagrangian model. The
expressions for the in-medium meson propagators, which take into account the
effects of matter polarization and renormalized vacuum polarization, are used
in the calculation of the various meson contributions to the total cross
section in the one-boson exchange approximation. The final expressions allow a
satisfactory fit of the Coulomb substracted proton-proton cross section data in
vacuum. At finite density or temperature the cross section is reduced as
compared to the vacuum values and compatible with the existing experimental
constraints. At high density the $p$-$p$ differential cross section gets 
increasingly forward and backward peaked. The $\sigma$-$\omega$ mixing 
plays a key role in the understanding of the results.

%\PACS{25.40.Cm,21.30.+y,21.65+f,24.10.Jv,21.60.Jz}

\newpage

{\bf (1) INTRODUCTION}

In plasma physics the screening effects of the medium modify the effective
interparticle interactions which often become very different than in vacuum.
The knowledge of the effective interaction is necessary for a proper analysis
of the collective behavior of the plasma. In the particular example of nuclear
matter, the screening of the medium on the N-N interaction (described in vacuum
by a two-nucleon potential which has a Yukawa-like expression), leads to a new
form of the effective potential which is long-ranged and oscillatory
\cite{DPS89,GDP94}. Such "Friedel-like" oscillations can induce, under
particular conditions, the appearance of some new observable properties in the
nuclear plasma, which should be the analogues of the ones encountered in
degenerate electromagnetic plasmas. 

Another important parameter related to the nuclear interaction, which can be
modified by the screening, is the N-N cross section. It is a basic ingredient
in the dynamical analysis of heavy ion collisions. Most of the early
calculations in this field have been performed using a constant value of the
cross section ($\sigma$ = 40 mb) or a fit of the experimental data in vacuum
\cite{CM81}. However, there is an ever growing bulk of evidence that the
modifications introduced by in-medium effects on the cross section have a
non-negligible influence on the findings of these calculations. For example, a
criterion which is often used to determine the stiffness of the equation of
state is the amount of transverse flow. But the collective flow, and its 
disappearance at the balance energy, is also used to set constraints on 
the magnitude of the N-N cross section in the medium. It is therefore not  
possible to disentangle completely the effects coming from the equation 
of state or from a reduced cross section. It has lately been found by several
authors, from BUU \cite{KW93,W93,MSFRS92,L93,AR95} as well as QMD \cite{KO92}
or AMD \cite{TOHME95} calculations, that a lower cross section in the medium 
is favoured by a detailed analysis of the transverse flow. Moreover, 
the cross section is an essential parameter in the calculation of transport 
coefficients and the knowledge of its screened behavior allows the 
determination of these coefficients as functions of the thermodynamical state. 

There exists several calculations \cite{F89} - \cite{GZ93} of the in-medium
elastic N-N cross section. Only one of these calculations considers finite 
temperature \cite{AR94} as well as density. Most of them were done in the 
Brueckner formalism and are appropriate for the low energy range $E_{lab} 
\simeq [50-300]$ MeV. Nevertheless, the relativistic effects should become 
important not only for higher energies, but (due to the medium effects) for 
high densities and temperatures. The purpose of this paper is to perform a 
fully relativistic calculation of the screening effects of the medium on 
the elastic N-N cross section in nuclear matter at finite density and 
temperature. 

An acceptable (phenomenological) description of the nuclear interaction in
terms of meson exchanges requires the analysis of several (scalar,
pseudoscalar, vector, pseudovector and tensor) meson couplings \cite{M89}. 
This work is a step in a program of analysis of the screening effects on 
the complete nuclear interaction in nuclear matter, where all these 
couplings will be considered. Here we treat only the case of the exchange 
of scalar ($\sigma$), pseudoscalar ($\pi$) and vector ($\omega$) meson 
couplings. Although the present treatment of the interaction is not 
complete, it allows for a good fit of the experimental data of the 
elastic N-N cross section \footnote{ Here, we restrict to the Coulomb 
substracted proton-proton p-p or neutron-neutron n-n cross section} 
in vacuum and gives a satisfactory description of the screening effects 
on this parameter. Indeed, preliminary calculations including the Yukawa 
and derivative tensor couplings to the $\rho$ meson, which will be 
published elsewhere, indicate that the corrections to the present 
calculations introduced by these suplementary couplings are small. 
Nevertheless, due to the isospin asymmetry associated to the $\rho$-N 
coupling, the proton-neutron (p-n) elastic cross section is more affected 
by the $\rho$-exchange contributions than the p-p and n-n ones. 

At highly relativistic energies, the $\Delta(1232)$ resonance and 
eventually more massive resonances such as $N^*(1440)$ are expected 
to play an important role. We calculated here the elastic channel 
of the N-N cross section and did not consider these resonances.
The way in which the taking into consideration of the $\Delta$ resonance 
might modify our analysis will be discussed in the last section. 
Basically two types of corrections will have to be considered:

First, the inelastic channels $N+N\rightarrow N+\Delta$ and $N + N
\rightarrow \Delta + \Delta$ open at $T_{lab} \ge$ 350 MeV. At energies 
higher than 800 MeV, the inelastic contribution represents 50\% of the 
total N-N cross section. The $\Delta$ production and absorption {\it via}
the inverse processes $N+\Delta\rightarrow N+N$ ... {\it etc} mechanism is 
the input for the calculation of the subsequent pion production through 
$\Delta$ decay ($\Delta \rightarrow N+\pi$). The theoretical pion 
distribution and spectra obtained in this way serve as a basis for 
comparison with the values observed experimentally in relativistic 
heavy ion collisions. This important topic deserves a separate study 
and will be treated elsewhere. 

Secondly, the $\Delta$'s can also appear in internal loops. They will
thus modify the screening of the interaction. There exists an abundant 
litterature on the modification of the $\pi$ dispersion relation by the 
$\Delta$ (see {\it e.g.} \cite{OTW82,HWB92,AH94} ). This topic is not 
yet completely settled, especially concerning the relativistic treatment. 
We argue that the screening of the pion will not play an important role 
for the proton-proton cross section calculated here, and will discuss the
case of the modifications to the $\sigma$ and $\omega$ dispersion
relations in the last section (see section (6)). As far as we know, the 
modification of other mesons than the $\pi$ by delta-hole loops has not yet 
been considered in the litterature; this will be the subject of future work.

In section (2) we give the lagrangian model, define the Hartree equilibrium and
introduce the in-medium meson propagators at finite temperature, obtained in
the one-loop approximation and including the vacuum polarization effects.
Section (3) is devoted to the calculation of the elastic N-N cross sections
from the screened propagators in the one-boson exchange approximation. In
section (4) we determine the model parameters in order to fit the available
experimental data on cross section in vacuum. In section (5) we present
the results with the analysis of the effects of density and temperature 
on this variable, and conclude in section (6) with a discussion
on the possible extensions of the model.

\vskip 1cm

{\bf (2) THE MODEL AND THE MESON DISPERSION RELATIONS}

\vskip 0.5cm

{\bf 2.1 Lagrangian density}

The cross section will be calculated by considering that the nucleons are
interacting through the exchange of $\sigma$, $\omega$ and $\pi$ mesons. The
dynamics of the model is given by the Lagrangian density 

\beq
\hat{L}\ & = & \ {i\over 2}\,[\, \hat{\overline{\psi}}\, \gamma \cdot 
              (\partial \hat{\psi})\, - \, (\partial\hat{\overline{\psi}})
              \cdot \gamma \hat{\psi}\,]\, -\, m \, \hat{\overline{\psi}} 
              \hat{\psi} \nonumber \\
         &   & +\ {1\over 2} \, [\,( \partial^{\,\nu} \hat{\sigma})
               ( \partial_{\,\nu} \hat{\sigma})\, -\, \mu_{\sigma}^{\ 2} 
               \hat{\sigma}^{2}\, ] \nonumber \\
         &   & +\ {1 \over 2} \,[\,( \partial^{\,\nu} \hat{\vec{\pi}})
               ( \partial_{\,\nu}\hat{\vec{\pi}})\, -\, \mu_{\pi}^{\ 2} 
               \hat{\vec{\pi}}^{2}\, ] \nonumber \\
         &   & - \ {1 \over 4} \, \hat{F}_{\omega}^{\,\mu\nu} 
               \hat{F}_{\omega\,\mu\nu}\,- \,
               {1 \over 2} \,\mu_{\omega}^{\ 2} \,\hat{\omega}
                ^{\,\nu} \,\hat{\omega}_{\nu}  \nonumber \\
         &   & \ +\ g_{\sigma} \,\hat{\overline{\psi}} \,\hat{\sigma} \,
               \hat{\psi} \,+\, g_{\omega} \,\hat{\overline{\psi}} 
               \,\gamma^{\mu} \,\hat{\omega}_{\mu} \,\hat{\psi}\,+\,  g_{\pi} 
               \,\hat{\overline{\psi}} \,\gamma^{5}\,\vec{\tau}\hat{\vec{\pi}} 
               \,\hat{\psi}  
\label{eq:(2.1)}
\eeq

where

\beq
\hat{F}_{\,\omega}^{\,\mu\,\nu}\ =\  \partial^{\,\mu} \hat{\omega}^{\,\nu}
\ -\ \partial^{\,\nu} \hat{\omega}^{\,\mu}
\eeq

and the hat symbol over characters denotes quantum operators.

A meson-meson interaction term of the form $\sigma . \vec{\pi}^{2}$ is usually 
added to this lagrangian model in order to improve the fit of the pion-nucleon 
scattering amplitudes in vacuum. If the value of the associated coupling 
constant is properly constrained, cancellations of large contributions to the 
s-wave scattering lengths occur and the predictions of the model become 
reasonable in vacuum \cite{SW86}. Nevertheless, no such cancellations occur 
at finite density and temperature, unless this coupling constant be differently
constrained in every thermodynamical state. Moreover, for some values of the
coupling strength this term can lead to difficulties related to the tachyonic
poles introduced in the pion propagator by the imaginary character of the
effective mass of the quasi-pion in the Hartree approximation. Finally, as we
have tested, this term has no important quantitative incidence on the
calculated cross sections in this approximation. Consequently, we shall not
consider this interaction here. 

In the calculations performed in this paper, the numerical values of the
constants appearing in the Lagrangian (\ref{eq:(2.1)}) are fixed as follows:
The meson and fermion masses are fixed to their "physical" values $
\mu_{\sigma} \ = \ 550 $ MeV, $ \mu_{\omega} \ = \ 783 $ MeV, $\mu_{\pi} \ = \
138 $ MeV and $ m = 939 $ MeV. The solution of the model in the relativistic
Hartree approximation (first order approximation in our scheme) leads to the
equation of state of the plasma \cite{CHIN77}. Then, we chose for the $\omega$
and $\sigma$ coupling constants the values which give a satisfactory fit of the
saturation properties in this approximation. These are $ g_{\sigma}^{2} \ = \
183.3\ (\mu_{\sigma}/m)^{2} $ and $g_{\omega}^{2} \ = \ 114.7\
(\mu_{\omega}/m)^{2} $. With these values, saturation is attained at $ P_{f0} =
1.40\ fm^{-1} $ with a binding energy $E_{b} = -15.85$ MeV. Nevertheless, in
the higher order approximations leading to the calculation of the in-medium
elastic cross sections we shall often use different values of the coupling
constants in order to fit the experimental values of this parameter in the
vacuum (see section 4). This is justified because in our cluster expansion
scheme \cite{DA85} the values of the parameters must be redefined at every
order. 

\vskip 0.5cm

{\bf 2.1 Dispersion relations}

Beyond the Hartree approximation, we treat the in-medium N-N interaction as
mediated by dressed mesons. The one-loop $\omega-\sigma$ meson propagators at
$T=0$ have been obtained for the first time in Ref. \cite{MS82}.  The
dispersion relations and propagators of the mesons at finite temperature used
here were obtained from a linear kinetic analysis of perturbations around the
Hartree ground state, starting from the relativistic quantum B.B.G.K.Y.
hierarchy \cite{H78,DP91}. This cluster expansion differs from the usual loop
expansion, but one can show that the first order of the B.B.G.K.Y expansion
coincides, at zero temperature, with the one-loop approximation \cite{GDP94}.
Nevertheless, the higher order corrections in the kinetic expansion differ from
the results of many-loop calculations.

Renormalized vacuum polarization contributions are crucial in obtaining a
physically reasonable behavior of the propagation modes \cite{DP91}. A first
analysis of the structure of the one-loop $\sigma-\omega$ propagators at zero
temperature has been performed by Lim and Horowitz \cite{LH89} in the
semiclassical approximation (no vacuum polarization effects). It has been
partially extended to the finite temperature case with vacuum effects in
\cite{SM89} and (completed with a study of the the zero-sound modes and a
improved choice of the renormalization point) in Ref. \cite{DFH92}.  

The dispersion relation matrix as obtained in Ref. \cite{DP91} has the form

%\beq
%D(k)\ = \ \left[
%\begin{array}{ccc}
%-k^{\mu}\,k^{\nu}\,+\,(\,k^{2}-\mu_{\omega}^{2}\,)\,g^{\mu\nu}\,
%-\,g_{\omega}^{2}\Pi^{\mu\nu}_{\omega}(k) &            
%g_{\sigma}g_{\omega}\Pi_{\sigma \omega}^{\mu}(k) &    0     \\
%          &              &            \\
%-g_{\sigma}g_{\omega}\Pi_{\sigma\omega}^{\nu}(k)   &    
%-k^{2}\,+\,\mu_{\sigma}^{2}\,-\,g_{\sigma}^{2}\,\Pi_{\sigma}(k) &  0  \\
%          &              &            \\
%    0     &        0     &      
%k^{2} \ - \ \mu_{\pi}^{2}\ + \ g_{\pi}^{2}\Pi_{\pi}(k)
%\end{array} \right] 
%\label{eq:(2.2)}
%\eeq

\beq
D(k)\ = \ \left[
\begin{array}{ccc}
-k^{\mu}\,k^{\nu}\,+\,(k^{2}-\mu_{\omega}^{2})\,g^{\mu\nu} &
g_{\sigma}g_{\omega}\Pi_{\sigma \omega}^{\mu}(k) &    0     \\
-\,g_{\omega}^{2}\Pi^{\mu\nu}_{\omega}(k) \phantom{\hat I} &       &  \\
          &              &            \\
-g_{\sigma}g_{\omega}\Pi_{\sigma\omega}^{\nu}(k)   &    
-k^{2}\,+\,\mu_{\sigma}^{2} -\,g_{\sigma}^{2}\,\Pi_{\sigma}(k) &  0  \\
          &              &            \\
    0     &        0     & k^{2} \ - \ \mu_{\pi}^{2} 
                           + \ g_{\pi}^{2}\Pi_{\pi}(k)
\end{array} \right] 
\label{eq:(2.2)}
\eeq

where the general expressions for the polarizations $\Pi(k)$ of the meson
fields as functions of the thermodynamical state (density and temperature) and
4-momentum $k^\mu=(\omega,\vec q)$ are summarized in appendix A. The $\pi$
propagation is decoupled from the dynamics of the other mesons because of
parity conservation. There is a mixing between the $\sigma$ and
longitudinal-$\omega$ propagation modes. The dynamics of the mesons in the
medium is affected by the real as well as the imaginary parts of the
polarizations. As mentioned above, the analysis of the propagation modes
associated to the real components of the polarizations has been performed
already. Let us make some comments about the behavior of the imaginary
polarizations (see also appendix A.2): 

a) Except for the mixing terms, the baryon and antibaryon contributions to the
imaginary parts of the polarizations have opposite signs. The antibaryon
contributions vanish in the $ T \ = \ 0 $ limit and the baryon contributions
reduce to the usual expression for the matter contribution to the imaginary
polarizations. The vacuum contributions to the imaginary parts have the same
form as in the $ T \ = \ 0 $ case and are finite. 

b) For time-like modes, there are two regions in the first sector of the
$q-\omega$ plane separated by the parabola defined by 

\beq
     \omega \  = \  \sqrt{q^2\ + \ 4M^2}  \qquad\qquad
     (\Delta \ = \  1 \ - \ 4 M^2 /(\omega^2 \ - \ q^2) \ = \ 0). 
     \label{eq:(2.3)}
\eeq
 
Here, $M=m-g_\sigma <\sigma>$ is the effective baryon mass. As in the $ T \ = \
0 $ case the imaginary parts vanish in the region between this parabola and the
diagonal $ \omega \ = \ q\ (\Delta < 0) $, and the modes are undamped there.
Over this parabola $ (\Delta > 0) $ the modes are damped by the decay into
particle-antiparticle pairs. 

c) For spacelike modes in the $ T \ = \ 0 $ limit, the imaginary parts vanish 
outside the region defined by

\beq
     \sqrt{(p_F - q)^2 \ + \ M^2} - \sqrt{p_F^2 \ + \ M^2} < \omega <
     \sqrt{(p_F \ + \ q)^2 \ + \ M^2} - \sqrt{p_F^2 \ + \ M^2}
     \label{eq:(2.4)}
\eeq

($p_F$ is the Fermi momentum) where the modes are damped by the decay into
particle-hole pairs. This region arises because, at $ T \ = \ 0 $, the baryonic
Fermi factor in Eq.(\ref{eq:(A.20)}) becomes a step function. At finite
temperature the imaginary parts are finite for any space-like mode.
Nevertheless, for moderate temperatures, the behavior of the Fermi factors
makes the imaginary parts small outside the particle-hole region. 

\vskip 1cm

{\bf (3) THE O.B.E. CROSS SECTION}

The in-medium cross section is calculated by describing the nucleon interaction
through the exchange of single dressed $\sigma$, $\omega$ and $\pi$ mesons. The
generic O.B.E. diagram is given in Fig.(1). The fermion lines correspond to the
nucleons in the medium which acquire an effective mass and momentum from the
ambient mean field in thermodynamic equilibrium. The properties of the
effective nucleons are calculated in the Hartree approximation with
renormalized vacuum fluctuations taken into account \cite{D85}. The dressed
propagator matrix of the mesons is obtained by the inversion of the dispersion
relation matrix (\ref{eq:(2.2)}). 

\vskip 0.5cm

{\bf 3.1 Dressed meson propagators}

The expressions of the polarizations given in appendix A were calculated, 
without loss of generality, in a frame (R1) where the background fluid is at 
rest and the momentum tranfer is $k^{\mu}=(\omega,0,0,q)$. In the following 
we will be working in the center of mass system of the collision (R2) in 
which the matter comoving system travels with the quadrivelocity $u^{\mu}$. 
It will therefore be more convenient to put the dispersion relations 
and propagators under a compact covariant form.

We define the auxiliary quadrivector $\eta^\mu$ and the projectors onto 
the longitudinal ($\Lambda^{\mu\nu}$) and transverse (${\cal T}^{\mu\nu}$) 
modes as follows:
\beq
\eta^{\mu} & = & u^{\mu} - {k.u \over k^2} k^{\mu} \\
\tilde g^{\mu\nu} & = & g^{\mu\nu} - {k^\mu k^\nu \over k^2} \\
{\cal T}^{\mu\nu} & = & \tilde g^{\mu\nu} - {\eta^{\mu} \eta^{\nu} \over 
\eta^2} \\
\Lambda^{\mu\nu} & = & {\eta^\mu\eta^\nu \over \eta^2}
\eeq
so that the polarizations now read
\beq
\Pi_{\sigma\omega}^\mu &=& \Pi_{\times} \eta^{\mu}/[\eta^2]^{1/2} \\
\Pi_\omega^{\mu\nu}&=&-\Pi_{\omega L} \Lambda^{\mu\nu}
-\Pi_{\omega T} {\cal T}^{\mu\nu}
\eeq
in terms of the longitudinal, transverse and mixing parts $\Pi_{\omega L}$, 
$\Pi_{\omega T}$ and $\Pi_{\times}$. These scalar quantities are related to 
the components $\Pi_\omega^{00}$, $\Pi_\omega^{11}$, $\Pi_{\sigma\omega}^{0}$ 
defined in  appendix A by
\beq
\Pi_{\omega L}={\omega^2-q^2 \over q^2} \Pi_{\omega}^{00} \quad ; \quad
\Pi_{\omega T}=\Pi_{\omega}^{11} \quad ; \quad
\Pi_{\times}^2={q^2-\omega^2 \over q^2} (\Pi_{\sigma\omega}^{0})^2 \nonumber
\eeq

The expression for meson propagator matrix becomes

\beq
   G^{mn}=\left(
   \begin{array}{ccc} G_{\omega}^{\mu\nu} & G_{\sigma\omega}^{\mu} & 0 \\ 
                      G_{\omega\sigma}^{\mu} & G_{\sigma} & 0 \\
                      0 & 0 & G_{\pi}  
   \end{array}\right)
   \label{eq:(3.1)}
\eeq

where the different elements (propagators) are given by

\beq
G_{\sigma}& =& i {k^2-\mu_{\omega}^2+g_\omega^2 \Pi_{\omega L}
   \over (k^2 -\mu_{\sigma}^2 + g_{\sigma}^2 \Pi_{\sigma})(k^2 -\mu_{\omega}^2 
   +g_\omega^2 \Pi_{\omega L} ) +g_\sigma g_\omega \Pi_{\times}^2} 
   \label{eq:(3.2)}
\eeq
\beq
G_{\omega}^{\mu\nu} & = & -i \Bigl[ {\cal T}^{\mu\nu}\ {1 \over 
   k^2 -\mu_{\omega}^2 +g_\omega^2 \Pi_{\omega T}} -{k^{\mu} k^{\nu}
   \over \mu_\omega^2 k^2}  \nonumber \\
& & + \Lambda^{\mu\nu}\ {k^2 -\mu_{\sigma}^2 + \ g_{\sigma}^2 \Pi_{\sigma}
   \over (k^2 -\mu_{\sigma}^2 + \ g_{\sigma}^2 \Pi_{\sigma})
   (k^2 -\mu_{\omega}^2 +g_\omega^2 \Pi_{\omega L}) +g_\sigma g_\omega 
   \Pi_{\times}^2} \Bigr] \\
& = & -i \left[ G_{\omega T} {\cal T}^{\mu\nu} +G_{\omega L} \Lambda^{\mu\nu}
       -{k^\mu k^\nu \over \mu_\omega^2 k^2} \right] \nonumber \\
   \label{eq:(3.3)}
\eeq
\beq
G_{\sigma\omega}^{\mu} & = & -i \eta^{\mu} {g_\sigma g\omega \ 
    \Pi_{\times} \over (k^2-\mu_{\sigma}^2 +\ g_{\sigma}^2 \Pi_{\sigma})
    (k^2 -\mu_{\omega}^2 +g_\omega^2 \Pi_{\omega L} ) +
    g_\sigma g\omega \ \Pi_{\times}^2} \\
& = & -i \eta^\mu G_\times \nonumber
   \label{eq:(3.4)}
\eeq
\beq
G_{\pi}& =& i {1 \over k^2 -\mu_{\pi}^2 + \ g_{\pi}^2 \Pi_{\pi}} 
   \label{eq:(3.5)}
\eeq

In inverting the dispersion matrix Eq.(\ref{eq:(2.2)}), the meson branches
associated to the zeroes of the determinant become poles of the meson
propagators. We are interested here in the spacelike region of the $ q - \omega
$ plane and, when vacuum polarization is taken into account, there are only
three poles of the propagator in this region, associated to the "tachyonic"
branches of the mixed $ \sigma $-longitudinal, transverse and pion
modes \cite{DP91}. These branches come from the vacuum polarization terms and
arise at large $k$, where the one-loop calculations fails \cite{FH88}. In fact,
for these values of $k$ (short distances) the point particle approach becomes
inadequate and the structural properties of nucleons must be taken into
account. In a more fundamental theory of pointlike particles, vertex 
corrections could be calculated self-consistently (see {\it e.g.} \cite{W95}). 
In nuclear matter models, such corrections have been introduced in analogy with
the QED case, with the suppression of high momentum transfer arising 
from the virtual Bremsstrahlung of neutral vector $\omega$ mesons 
\cite{KNPW93,AS92-ST95}. In our phenomenological model, this can as well 
be done through the introduction of phenomenological monopolar form factors 
for the nucleon at every vertex in the boson exchange
diagrams as well as at every vertex in the loop calculation of the boson
propagators. This leads to new expressions which, in $k$-space, are obtained
from the original ones by multiplying every coupling constant by the
corresponding form factor. These form factors have the generic form 

\beq
   {\cal F}(k)\ =\ (\Lambda^{2} \ - \ \mu^{2})/(\Lambda^{2} \ - \ k^{2})
   \label{eq:(3.8)}
\eeq   

where $ \mu $ is the scalar, vector or pion meson mass in every case and $
\Lambda $ is a cut-off parameter. After the introduction of the form factors,
if the cut-offs are in appropriate ranges, the spurious "tachyonic branches" as
well as the associated poles of the meson propagators on the $ \omega < q $
region disappear. The renormalized vacuum contributions are crucial in
obtaining physical (stable and causal) propagation modes \cite{DP91}. If they
are neglected, new unphysical spacelike modes are present at small $q$ which 
can not be eliminated by the form factors and adulterate the pole structure 
of the propagators. 

\vskip 0.5cm

{\bf 3.2 Transition matrix and N-N cross section}

The O.B.E. amplitude ${\cal M}$ is obtained in the first Born approximation
from the diagram of Fig.(1). In condensed form, we have

\beq
   {\cal M}=\sum_{m,n=\sigma,\omega,\pi} (\overline U_3 \Gamma_m U_1) 
   G_{31}^{m n} 
   (\overline U_4 \Gamma_n U_2) - \hbox{\rm exchange} 
   \label{eq:(3.10)}
\eeq

In this expression the $U$'s are the spinors defining the incoming 
($ U_1, U_2 $) and outgoing ($ U_3, U_4 $) nucleon states. The indices
$i=1,2,3,4$ stand for the effective momenta $(p_i^\mu)^*=p_i^\mu-g_\omega
<\omega^\mu>$ and the spin-isospin quantum numbers $s_i$, $t_i$.
$G_{31}^{mn}$ is the propagator matrix (\ref{eq:(3.1)}) which depends
on the momentum transfer $k_{31}^\mu=p_1^{\mu\, *}-p_3^{\mu\, *}$.
$ \Gamma $ is the coupling matrix defined by 

\beq
   \Gamma = \left(
   \begin{array}{c}  -i g_{\omega} \gamma^{\mu} \\
                     i g_{\sigma} \\ 
                      g_{\pi} \gamma_5 \vec \tau
   \end{array}
   \right)
   \label{eq:(3.11)}
\eeq

In the explicit calculation of the transition matrix from these formulae, all
the coupling constants in $ G^{mn} $ and $ \Gamma_m $ are to be multiplied by
the corresponding form factors. The exchange term is obtained by interverting
the indices 3 and 4. 

In the referential (R1) where the fluid is at rest, $u^{\mu} =(1,\vec 0)$ and
\beq 
p_1^{\mu} = P^{\mu} + k^{\mu}/2 \quad ; \quad 
p_2^{\mu} = K^{\mu} - k^{\mu}/2 \quad ; \quad 
p_3^{\mu} = P^{\mu} - k^{\mu}/2 \quad ; \quad 
p_4^{\mu} = K^{\mu} + k^{\mu}/2 \nonumber
\eeq

with

\beq
k^{\mu} = (\omega,0,0,q) \nonumber 
\eeq
In the center of mass of the collision (referential (R2)), the background 
fluid moves with a velocity $u^{\mu}=(\gamma, \gamma \vec v)$
parametrized by
\beq
\vec v = \left(\matrix{ v \sin \alpha \sin \varphi \cr 
                        v \sin \alpha \cos \varphi \cr 
                        v \cos \alpha \cr} \right)
\eeq
The momenta of the incident and scattered particles are (energy conservation
is already taken into account)
\beq
p_1^{\mu} = (E,\vec{p}) \quad , \quad 
p_2^{\mu} = (E,- \vec{p}) \quad , \quad 
p_3^{\mu} = (E,\vec{p}\, ') \quad , \quad 
p_4^{\mu} = (E,- \vec{p}\, ') \quad , \nonumber
\eeq

the scattering angle is $\theta$ defined by 
$\vec p.\vec{p}\, '/ \sqrt(\vec{p}^{\, 2} \vec{p}^{\, '\, 2}) = \cos \theta$.
The momentum transfers for the direct ($k_{31}$) and exchange ($k_{41}$) 
diagrams are

\beq
k_{31}^{\mu} &=& (0,\vec p -\vec{p}\, ') \nonumber \\
k_{41}^{\mu} &=& (0,\vec p +\vec{p}\, ') \nonumber 
\eeq

One goes from (R1) to (R2) by performing a boost of four-velocity
$(\gamma, \gamma \vec v)$ and a rotation.
In the referential (R1), 
\beq
k_{31}^\mu (R1) & = &  (\omega_\ominus,\vec q_\ominus) \nonumber  \\
k_{41}^\mu (R1) & = &  (\omega_\oplus,\vec q_\oplus) \nonumber \\
\omega_{({\ominus \atop \oplus})} & = & \gamma v p [\pm \sin \alpha 
\sin \varphi \sin \theta - \cos \alpha (1 \mp \cos \theta)] \nonumber \\
q_{({\ominus \atop \oplus})} & = & [\omega_{({\ominus \atop \oplus})}^2 +
2 p^2 (1 \mp \cos \theta) ]^{1/2} \nonumber 
\eeq

The expression of the differential cross section for elastic N-N scattering 
in the center of mass of the two colliding particles is given by 

\beq
   d \sigma / d \Omega = 1/(64 \pi^2 s) |{\cal M}|^2
   \label{eq:(3.9)}
\eeq

where $s$ is the Mandelstam variable $s=4 E^2$. 
In the following we consider only the spin-averaged cross section 
obtained by replacing $|{\cal M}|^2$ by
$\overline{|{\cal M}|^2}=(1/4) \sum_{s_1,s_2,s_3,s_4} |{\cal M}|^2$ in 
Eq. (\ref{eq:(3.9)}).
We obtain the $p$-$p$ cross-section for the isospin quantum numbers 
$t_1=t_2=t_3=t_4=1$ and the $n$-$n$ cross section for $t_1=t_2=t_3=t_4=0$. 
$\sigma_{pp}$ and $\sigma_{nn}$ are equal in our approximation where 
the Coulomb interaction was substracted and the neutron and proton masses 
taken to be equal.

The explicit formula for $d \sigma/d \Omega$ is given in Appendix B in terms
of the Mandelstam variables. 

In the next section the values of the coupling constants and the cut-offs of
the form factors will be fixed in order to fit the cross section experimental
data in vacuum. 

\vskip 1cm

{\bf (4) CHOICE OF MODEL PARAMETERS}

\vskip 0.5cm

{\bf 4.1 Vacuum effects}

We recall that the vacuum effects were renormalized (see appendix A)
in such a way that the contribution of the vacuum polarizations at vanishing
temperature and density cancel on the mass shell of the mesons:
\beq
\Pi^{vac}_{i}(k=k_i) =0 \ \mbox{\rm at}\ k_i^2=\mu_i^2,\
i=\{\sigma,\ \omega,\ \pi \}
\eeq
For other values of $k$ however, the polarisation of the vacuum will bring a
contribution to the free propagator
\beq
G^0_i(k)={1 \over k^2-\mu_{i*}^2(k)} \qquad , & \qquad 
 \mu_{i*}^2(k)=\mu_i^2 -g_i^2(k) \Pi^{vac}_i(k) & \nonumber \\
& {\rm with}\ g_i(k)=g_i.\displaystyle{\Lambda_i^2-\mu_i^2 \over 
\Lambda_i^2 - k^2} &
\eeq
The taking into account of the vacuum polarization has the effect of
reducing the effective $k-$dependent meson masses $\mu_{i*}(k)$
at high momentum transfer $k$. As mentioned above, the vacuum polarization 
contributions are essential in obtaining a reasonable pole structure of the
propagator \cite{DP91}).

\vskip 0.5cm

{\bf 4.2 Fitting procedure}

Our model includes only $\pi$, $\sigma$, $\omega$ meson exchange. It is a
minimal model which allows to reproduce the general features of the
nucleon-nucleon cross section with only five parameters: the couplings of the
scalar and vector mesons $g_\sigma$, $g_\omega$ and the cutoffs
$\Lambda_\sigma$, $\Lambda_\omega$, $\Lambda_\pi$. We fix the 
pion-nucleon coupling
constant from experimental data on pion-nucleon scattering. Recent
work (see {\it e.g.} \cite{Arndt}) quote a value slightly lower
($g_\pi^2/(4\pi)=13.7$) than the standard textbook value
($g_\pi^2/(4\pi)=14.4$). We chose the latter one in order to be consistent with
previous work. This will have no important incidence on our results anyway. The
meson masses are kept fixed at their physical value $\mu_\pi=138$ MeV and
$\mu_\omega=783$ MeV. The scalar meson is not physical but is a
phenomenological way of describing correlated two pion exchange, so that the
constraint on its mass is less important \footnote{The $\sigma$ might 
be identified with the $f_0$ (400-1200) broad resonance mentioned in the
last issue of the Particle Data Book \cite{PDG96}. The controversy about
the identification of this scalar particle as a broad $\pi-\pi$ resonance
was revived by recent observations \cite{S96-T96}}. We take the standard 
Bonn potential value $\mu_\sigma=550$ MeV. The value of the $\omega$-$N$-$N$
coupling constant is not very well known experimentally. This situation
may be improved by future planned experiments near the $\omega$ production
threshold. Quark models \cite{HO84,KVWM88,ST94} predict $g_\omega \simeq 
8.5  - 8.8$. From the decay width $\Gamma(\omega\rightarrow e^+e^-)$ and in 
the vector dominance approximation, a value $g_\omega 
\simeq 8.6$ can be estimated. On the other hand, the Bonn model value
$g_\omega^2/(4\pi)\simeq 20.$, obtained by fitting the nucleon scattering 
phase shifts is much larger, but it can be reduced by taking into account 
correlated $\pi-\rho$ exchange. We will let $g_\omega$ vary and will 
see in the next subsection that our fit yields values in the range 8-9.
The cutoffs should be chosen larger than the mass of the meson exchanged.
There is also an upper limit set by the constraint that the tachyonic modes
of the dispersion relations should be removed (see the discussion of section 
(3.1) and Ref. \cite{DP91}). In practice, we restrict the search to the
range $1.5\ \mu_i < \Lambda_i < 1700$ MeV.

Publication of experimental data is scattered throughout the litterature.
Recent compilations and smooth fitting of the available data are to be found in
Refs. \cite{nijmdata,saiddata}. These compilations do not substract the Coulomb
scattering in $p$-$p$ data at low momentum exchange. Coulomb scattering is most
important at low energy $T_{lab} < 40$ MeV and/or scattering angles $\theta <
10^\circ-20^\circ$. We consider here a model of the nuclear interaction only,
since the cross section we are calculating is intended for use in the current
models of nuclear matter (BUU, QMD \ldots), where the Coulomb interaction can
be neglected. Simple fits of Coulomb-substracted experimental data are given by
Cugnon {\it et al.} \cite{CM81,CL88} for use in their cascade model. Some
Coulomb-subtracted data points are also given in the older paper of Chen {\it
et al.} \cite{Chendata}. 

We performed a simultaneous fit of the total elastic cross-section
{\it and} the differential cross section at 9 distinct values of the 
c.m. energy. The fit was obtained using a weighed least-squares method
where the data points are chosen as follows: For the total cross section 
$\sigma_{pp}$, we used 12 points distributed over the energy range 
$T_{lab} \in [40-1000]$ MeV. For the differential cross section,
$d \sigma_{pp}/d \Omega$, we used 8 points distributed in the range 
of scattering angles $\theta \in [10^\circ-90^\circ]$ for each value 
of the energy $T_{lab}=$60, 100, 250, 350, 460, 560, 660, 800, 1000 MeV.
Fitting only $d \sigma_{pp}/d \Omega$ would in principle be enough since 
$\sigma_{pp}$ is obtained by integrating over $\theta$, but we obtain 
better results by taking $\sigma_{pp}$ -data with a weight of 10 \% -- 20 \%
into account. Very-low or very-high energy data for $d \sigma_{pp}/d \Omega$
was also attributed a lower weight. We found in this way $\chi^2$ values 
in the range 10 -- 12. 

Due to the reduced number of free parameters, our model is not be expected 
to do as accurate a job as the last Bonn \cite{M89,Bonn,BonnCD}, Nijmegen 
\cite{Nijmegen} or Argonne \cite{Argonne} potentials for example. 
Nevertheless it will be seen in the remaining that a reasonably good fit 
can be obtained for the $p$-$p$ (and, which is equivalent here, $n$-$n$) 
total and differential elastic cross section. It is more
difficult to reproduce the $p$-$n$ cross sections. As a matter of fact, the
$T=1$ isospin channel which is necessary for the calculation of $p$-$n$
scattering requires the presence of the $\rho$-meson exchange term. For this
reason, in this paper we show only results for the $p$-$p$ cross-section.
Inclusion of more mesons ($\rho$, $\eta$, $\delta$) and calculation of the
$p$-$n$ cross section is currently under way. 

\vskip 0.5cm

{\bf 4.3 Fitting parameters}

Depending on the details of the fitting procedure, we found several 
equivalent parameter sets. 

\begin{table}[htbp]
\begin{center}
\begin{tabular}{|c|c|c|c|c|c|}
\hline
 & $g_\sigma$ & $g_\omega$ & $\Lambda_{\sigma}$ & $\Lambda_\omega$
 & $\Lambda_\pi$ \\
\hline
set A & 3.80 & 9.31 & 1298.8 & 1240.5 & 362.1 \\
set B & 3.27 & 8.38 & 1622.0 & 1324.7 & 355.2  \\
set C & 4.87 & 8.99 &  951.3 & 1255.3 & 397.4  \\
\hline
set D & 7.93 & 8.93 & 703.3 & 1237.3 & 376.1 \\
\hline
set E & 8.9 & 37.7 & 1643.2 & 903.2 & 587.2  \\
\hline
\hline
 & $m_\sigma$ & $g_\omega$ & $\Lambda_{\sigma}$ & $\Lambda_\omega$ 
 & $\Lambda_\pi$ \\
\hline
set F & 763.3 & 8.06 & 912.2 & 1348.5 & 390.2 \\
\hline
\end{tabular}
\end{center}
\end{table}

The total elastic $p-p$ cross section $\sigma_{pp}$ is well reproduced in 
the energy range $[40 - 1000]$ MeV by the parameter sets (A,B,C). (see
Fig. 2). We also quote the somewhat extreme choices (D,E,F). By comparing 
our results at finite density and temperature, this will allow to investigate 
the sensitivity of our conclusions to the choice of model parameters. (D) is 
obtained by fixing the coupling constants to their saturation values (see 
section (2)), so that it contains only the three cutoffs as free parameters.
(F) is obtained by fixing the coupling constant of the $\sigma$ meson
equal to that of the pion and instead let vary the $\sigma$ mass.
Sets (A,B,C,D,F) all correspond to the same minimum in parameter space.
Sets (D) and (F) have to pay for a too high $g_\sigma$ by a correspondingly
low $\Lambda_\sigma$. Set (E) corresponds to another (not very stable)
minimum in parameter space. It is somewhat better at low energy
but grows rapidly bad at energies higher than 500 MeV. There a very strong
vector coupling is castigated by a small $\Lambda_\omega$. The best choice 
is set (A), then we have (B) $>$ (C) $>$ (D) $>$ (F) $>$ (E). In all cases 
we notice  that the cut-off parameter of the pion is smaller than usually 
admitted in Bonn parametrizations. This is due to the absence of compensation 
by the $\rho$ meson. Preliminary results indeed show that the inclusion
of the $\rho$ meson allows higher values of $\Lambda_\pi$. Still the
$\Lambda_\pi$'s of all sets shown here are always above twice and a half
the pion mass.

The differential cross section is somewhat more difficult to fit accurately 
(see Fig. 3). It is possible to perform a good quantitative fit at 
high energies. At lower energies, the fit becomes qualitative.
The same remarks about the goodness (badness) of the parameter sets
as for the total cross section apply here. The low energy behavior
would be improved by increasing $g_\sigma$.

If we look at the contributions of each meson separately, we see that the
value of the cross section is essentially driven by the $\sigma$ and $\omega$
mesons. In a way similar to what happens for the nuclear potential \cite{DP91},
the value of $\sigma_{pp}$ results from a delicate balance of $\sigma$ and
$\omega$ contributions. The pion describes the low momentum-transfer,
long-range part and the detail of the differential cross section at forward
angles. 

\vskip 1cm

{\bf (5) RESULTS AND DISCUSSION}

As a first important statement, it is found that the various choices of fitting
parameters essentially give the same behavior at finite density and
temperature. We display here results for the set of parameters (A), and comment
about the other sets when appropriate. 

\vskip 0.5cm 

{\bf 5.1 In-medium cross section from screening}

The total elastic $p$-$p$ cross section at zero temperature is displayed 
in Fig. 4 for the parameter set (A) and for various values of the density. 
It is seen that it decreases with increasing density. The reduction is 
larger at smaller momentum transfer and energy, so that the total cross
section becomes approximately constant throughout the whole energy 
interval for $n \simeq n_{sat}$. Further increasing the density does
not much modify $\sigma_{pp}$. 

We do not observe significant differences for the
total $p$-$p$ cross section calculated with parameters sets (A,B,C,D,F).
For sets (A,B,C), $\sigma_{pp}\ (n \ge 2\ n_{sat})$ settles around
$\simeq$ 17 mb for all $T_{lab}$. For sets (D,F), we have a somewhat
lower value $\sigma_{pp}\ (n \ge 2\ n_{sat})$ $\simeq$ 15.5 mb, but
we saw that both models had a too high scalar coupling $g_\sigma$ 
and a too low scalar $\Lambda_\sigma$ cutoff. The general features
(low-density reduction, flattening to a $T_{lab}$-independent value,
constant value at high density) are also obtained with set (E), but
it settles to a higher value $\sigma_{pp}\ (n \ge 2\ n_{sat})$ $\simeq$ 
71 mb. This high value is to be traced to the unreasonably strong
$g_\omega$ and low cutoff $\Lambda_\omega$.

We also calculated the total $p$-$p$ cross section at finite temperature.
The ratio of the value of the cross section in the medium to its value
in the vacuum is shown in Fig. 5 as a function of density $n/n_{sat}$
for seven values of the temperature $T$=0, 40, 80, 120, 160, 200 and 240 MeV 
and for two values of the energy of the beam energy $T_{lab} = 100$ MeV
and $T_{lab}$ = 300 MeV. This same ratio is displayed as a function of
temperature on Fig. 6a  for five values of the density $n/n_{sat}$ 
= 0, 0.1, 1., 2. and 4. for the same values of the beam energy. 
These results are summarized in Fig. 7 as a contour plot in the 
density-temperature plane for incident particle energies $T_{lab}$ = 
100, 300, 500 MeV.

At $n=0$, a similar behavior is obtained with increasing the temperature 
as at $T=0$ with increasing density (the $\sigma_{pp}$ 
cross section decreases at high $T$) but the temperature dependence 
is not as strong as the density dependence: There is almost no difference
between the $T=0$ and $T=120$ curves. The same quantitative
effect is obtained by heating the vacuum at $T$ = (150, 170, 200) MeV,
or by compressing cold matter at $n$ = (0.1, 0.5, 1) times the
saturation density. The cross section at $n=0$ becomes flat over
the energy range $[40 - 1000]$ MeV at $T$=250 MeV.

At fixed non vanishing temperature, the cross section is a 
nonmonotonous function of the density as can be seen on Fig. 5.
This figure also reveals that the cross section is more
sensitive to temperature in the density range of interest for
nuclear collisions $n/n_{sat}$= 0.2 -- 2.

At fixed non-vanishing density, the cross section is non monotonous
as a function of the temperature. It first slightly increases,
reaches a maximum and decreases rather rapidly at high $T$. The
turnpoint temperature increases with increasing density. In brief,
the behavior of the cross section $\sigma_{pp}$ with varying $T$ 
qualitativelly follows that of the nucleon effective mass (compare
Figs. 6a and 6b).

The remarks done on the example of parameter set (A) as to the behavior
of the total cross section with temperature hold for (B,C,D,F) parameter 
sets as well, save some unsignificant quantitative differences.

Let us finally discuss the density and temperature dependence of the 
differential $p$-$p$ cross section (see Fig. 8 and Fig 3c). Since 
$d \sigma_{pp}/d \Omega$ is quantitatively not so well reproduced 
as the total one, we have to be more cautious in our conclusions. 
We nevertheless notice some general features: at high density or 
temperature, the differential $p$-$p$ cross section becomes very 
forward peaked (or equivalently backward peaked since the $p$-$p$ 
cross section is symmetric around $\theta=90^o$). 
This trend is not affected by the same ``saturation'' effect as for the 
integrated cross section. At low density on the other hand, forward 
scattering is slightly suppressed with respect to the free value.

The behavior of the cross section as a function of density and temperature
can directly be related to that of the polarizations. As a matter of fact, 
the polarizations appreciably differ from zero when the momentum transferred
by the meson is lower than the threshold value $\sqrt{-k^2}=2\ p_F$ for 
particle-hole production/recombination \footnote{The condition 
$\sqrt{-k^2}=2\ p_F$ is the same as that leading to the Kohn singularity 
at the threshold for particle-hole production which was at the origin 
of the Friedel oscillations in the $N$-$N$ potential \cite{DPS89}.}.
When this occurs, the mixing between the sigma meson and the longitudinal 
mode of the omega meson through particle-hole loops is important, 
and the balance between the attractive and repulsive parts of the 
interaction is disturbed.

Let us recast the propagators of the mesons inside the medium 
in a form as close as possible to the free one ({\it e.g. } $G_\sigma^0=
1/(k^2-[\mu_\sigma^{\rm vac}(k)]^2)$):
\beq
G_\sigma &=& {K_{\rm mix} \over k^2-[\mu_\sigma^{\rm eff}(k)]^2+
i \mu_\sigma^{\rm eff}(k) \Gamma_\sigma(k)}  \\
G_\omega &=&  {K_{\rm mix} \over k^2-[\mu_{\omega L}^{\rm eff}(k)]^2+
i \mu_{\omega L}^{\rm eff}(k) \Gamma_{\omega L}(k)} \Lambda^{\mu\nu} +
 {1 \over k^2-[\mu_{\omega T}^{\rm eff}(k)]^2+
i \mu_{\omega T}^{\rm eff}(k) \Gamma_{\omega T}(k)} {\cal T}^{\mu\nu}
\eeq
with
\beq
[\mu_\sigma^{\rm eff}(k)]^2 &=& \mu_\sigma^2-g_\sigma^2 {\cal R}e\  
\Pi_\sigma(n,T,k)  \\
1/\Gamma_\sigma &=& \mu_\sigma^{\rm eff}/g_\sigma^2 {\cal I}m\  
\Pi_\sigma(n,T,k) 
\eeq
(and similarly for $\pi$, $\omega L$ and $\omega T$), and
\beq
K_{\rm mix} &=& {(k^2-\mu_\omega^2+g_\omega^2 \Pi_{\omega L})
(k^2-\mu_\sigma^2+g_\sigma^2 \Pi_{\sigma}) \over 
(k^2-\mu_\omega^2+g_\omega^2 \Pi_{\omega L})(k^2-\mu_\sigma^2
+g_\sigma^2 \Pi_{\sigma})+g_\sigma g_\omega \Pi_\times^2} 
\eeq

This defines momentum dependent ``effective masses'' and widths, and
a mixing parameter. The ratio of the ``effective masses'' to their 
vacuum values and the mixing parameter are shown on Fig. 9 for $T=0$ 
and $n/n_{sat}=0.1$. Also shown on this figure are the dynamical masses,
defined as the solution of the dispersion relation at zero 3-momentum
$k=(\omega_i^{\rm pl},\vec 0)$: $\mu_i^{\rm dyn}=\omega_i^{\rm pl}$, $D(k)=0$.
It can be seen that the mixing parameter deviates strongly from its
vacuum value ( = 1) as long as the momentum transfer is lower than 2 $p_F$.
The transverse part of the $\omega$ propagator remains almost unchanged.
On the other hand, the ``effective masses'' of the $\sigma$ and longitudinal 
part of the $\omega$ follow the behavior of $K_{\rm mix}$, the effective 
$\sigma$ mass being reduced and the effective longitudinal $\omega$ mass 
enhanced in the same proportion. 

The mixing parameter comes as a multiplicative factor in $G_\sigma$
and $G_{\omega L}$, so that the mixing could be absorbed in the redefinition
of effective couplings for the $\sigma$ and the longitudinal $\omega$, and 
be viewed as a weakening of the strength of these effective interactions.

Finally, the non-vanishing imaginary part of the polarizations 
which describes damping of the modes through particle-hole decay,
contributes a positive term in the denominator of the square of the
propagators and further reduces the value of the cross section in the medium.

It is now easy to understand the behavior of the differential cross-section 
at finite density. Since the momentum transfer is given in terms of the 
incident energy $E=T_{lab}+M$ and scattering angle $\theta$  by 
$k^2=-2\ (E^2-M^2)(1 \pm
\cos\theta)$, the condition $\sqrt{-k^2}<2\ p_F$ yields at constant energy the
$\theta$ range affected by medium effects: 

\beq
|\cos\theta| > 1-{2 p_F^2 \over E^2-M^2}
\eeq
This corresponds to the range where $d \sigma_{pp}/d\Omega$ becomes 
very forward peaked (and also backwards-peaked, since $d \sigma_{pp}/d\Omega$
is symmetric by $\theta \rightarrow \pi-\theta$).

\vskip 0.5cm

{\bf 5.2 Comparison with Brueckner calculations}

The calculation we performed here amounts to consider loop contributions, 
while the Brueckner calculations of Li and Machleidt \cite{LM94},
Faessler {\it et al.} \cite{F89}, Alm {\it et al.} \cite{AR94}
consider ladder contributions. In fact, both types of contributions
should be included. This is however still somewhat beyond the present 
state of the art, as a naive ladder-resummation of our screened propagators 
would lead to inconsistencies (see {\it e.g.} a discussion in \cite{JW94}).

Still it is interesting to study separately the effect of loops or ladders.
The Brueckner resummation is expected to be more appropriate for low
densities, since it is based on an expansion in $n a^3$ ($a$ stands for 
the range of the interaction). On the other hand, our loop resummation 
is expected to be better at higher densities \cite{A82} and energies. 
In the case of the Brueckner 
resummation, the corrections due to finite density and temperature come in as
Fermi factors on the intermediate (real) nucleon states on the uprights of the
ladders, while in the case of loop resummations, the Fermi factors are applied
on the intermediate (virtual) nucleon states forming the particle-hole loop.
Both models, although they consider very different physical effects, lead to
very similar results: decreasing $\sigma_{pp}$ with increasing density and
saturation for $n > 2\ n_{sat}$, flattening over the energy range. 

There are also some differences: we observe a modification of the
in-medium cross section at high energies too, although the effect
is smaller than at low energy. 
We also have a more rapid decrease with density for low values of
$n/n_{sat}$. This result should probably not be taken too seriously:
First, because, as already pointed out above, our model is not
optimized for this range, and second, because these densities at $T=0$
fall within the domain of spinodal $dP/d\rho <0$ instability.
Indeed, such a rapid variation of the cross section with the
density is not observed when the background matter is heated moderately 
($T \sim 20$ MeV) so as to come out of the unphysical spinodal region.

\vskip 0.5cm

{\bf 5.3 Comparison with experimental constraints}

During the last five years, it has become a general trend to investigate
the influence of a modified $N$-$N$ cross section in the medium in
numerical simulations in heavy ion collisions. We mention here, as a 
non-exhaustive sample, the work of Klakow {\it et al.} \cite{KW93}
or Alm {\it et al.} \cite{AR95} on BUU codes, of Khoa {\it et al.} 
\cite{KO92} on a QMD code, or of Tanaka {\it et al.} \cite{TH96} 
with the antisymmetrized AMD code. 

All these determinations are strongly model-dependent, but the fact
that they all predict roughly the same effect, {\it i.e.} a 20 \%
reduction of the in-medium cross section at $n\, =\, n_{sat}$,
although they rely on a very different physical analysis of the 
processes involved, sounds very encouraging. 

We compare in Fig. 10 our results as well as the Brueckner results of Li 
and Machleidt \cite{LM94} to the BUU prediction of Klakow {\it et al.} 
\cite{KW93} as well as the AMD prediction of Tanaka {\it et al.} \cite{TH96}.

Tanaka {\it et al.} effected simulations of $p$+nucleus collisions using 
their antisymmetrized version of quantum molecular dynamics (AMD), and 
extracted a reduction factor for the in-medium $N$--$N$ cross section 
with respect to its free value by comparing the results of the simulation 
to available experimental data in the energy range [0 - 200] MeV.
The $p$+${}^9$Be collision probes low densities in the surface, while
the $p$+${}^{12}$C, $p$+${}^{27}$Al, $p$+${}^{40}$Ca, which all give
similar results, are thought to probe the nuclear density $n=n_{sat}$.
The result of Tanaka {\it et al.} is represented on Fig. 10 by triangles
for $p+{}^{40}$Ca and diamonds for $p+{}^9$ Be.

Klakow {\it et al.} estimated the medium-modified $N-N$ cross section
by performing simulations of heavy-ion collisions $A+A$ with a 
Boltzmann-Uehling-Uhlenbeck transport model. The heavy ion collision 
takes place with an initial energy per nucleon 200 MeV and the system
is evolved until freeze-out. The information on the cross section 
is obtained from an analysis of the dependence of nuclear transverse 
collective flow on beam energy, and the measurement of the so-called 
balance energy $E_{bal}$ at which the flow disappears. This energy 
indicates a transition in which one goes from attractive to repulsive 
scattering. The authors assumed a density dependence $\sigma_{med}/
\sigma_{free}=1-\alpha (n/n_{sat})$ {\it i.e.} by taking the first 
order of a Taylor expansion at constant beam energy and zero temperature. 
The best agreement with experimental data was found to correspond to 
$\alpha=0.2$. The result of Klakow {\it et al.} is shown on Fig. 10 
by two arrows for densities $n=0.5\ n_{sat}$ and $n=n_{sat}$.

The Brueckner calculation of Li and Machleidt \cite{LM94} was described 
in the preceding paragraph. We display in Fig. 10 their result for the
$p$-$p$ cross section at $n=0.5\ n_{sat}$ (dotted line) and $n=n_{sat}$
(continuous line).

Our results for the in-medium $p$-$p$ cross section are represented on 
Fig. 10 by four dashed and dot-dashed lines at densities $n=0.1 n_{sat}$ 
and $n=n_{sat}$ and for temperatures $T=0$ and $T=50$ MeV, as limiting 
values of what density and temperature conditions can be reached in the 
collisions considered by Klakow {\it et al.} or Tanaka {\it et al.}.

We can see that there is a fair overall agreement between these four
approaches. Experimentally as well as theoretically, the cross section
is found to be reduced with respect to its free value. There is also
a qualitative agreement in the behavior with the beam energy: as the
energy increases, the reduction factor decreases and saturates to a 
constant value. Both theoretical approaches (Brueckner, ours) overpredict the
reduction of the $N$-$N$ cross section with respect to the experimental
estimates. Two reasons may explain the observed discrepancies: 
\par\noindent {\it (i)} The
theoretical approaches consider that the collision is happening inside infinite
nuclear matter, while in the experimental setup and the numerical simulations
finite size effects play an important role, so that in a dynamical situation
the dressing of the mesons may not be fully effective. 
\par\noindent {\it (ii)} The
calculation of Li and Machleidt as well as ours assume that the background
matter is not far from local equilibrium, while the initial state of the
collision is strongly out of equilibrium, which also modifies the dressing
of the mesons.
\par\noindent It is interesting to note that the discrepancies observed
with the T=0 calculations (Brueckner and ours) are reduced when finite 
temperatures are considered. This can be an indication that such finite 
temperature effects play a role in moderating the reduction of the cross 
section.

\vskip 1cm

{\bf (6) CONCLUSION AND OUTLOOK}

\vskip 0.5cm

{\bf 6.1 Concluding summary}

We have analyzed in this work some aspects of the screening effects on the
nucleon-nucleon interaction, as due to the modifications introduced by the
nuclear medium (at zero and finite temperature) on the propagation of the
mesons which mediate this interaction. We applied here our results on dressed
meson propagators to the calculation of the modification of proton-proton (or
neutron-neutron) cross section for collisions occuring inside dense and hot
nuclear matter. We found that the $p$-$p$ or $n$-$n$ total elastic cross
sections were reduced with increasing density. The effect is more important at
low energy, leading to a nearly constant cross section $\sigma_{pp} \simeq 17$
mb at all energies for densities above the saturation value. The temperature
dependence was found to be less pronounced, with first a slight increase and
then a reduction with respect to the free value. The differential cross section
becomes increasingly forward (and backward) peaked as the density or 
temperature grows. 

The general features of the behavior with density are found to be the same as
from a calculation with the Brueckner formalism effected by Li and Machleidt
\cite{LM94}, although both approaches are very different. The reduction with
density and its behavior with energy are compatible, although somewhat
overestimated, with respect to the information extracted from BUU and AMD
analysis of experimental data. Taking into account the effects of finite
temperature in our approach seems to improve the agreement. 

\vskip 0.5cm

{\bf 6.2 Improved meson exchange picture}

While the calculation presented in this paper was restricted to the
nucleon-nucleon scattering between identical particles through the exchange of
$\pi$, $\sigma$ and $\omega$ mesons, it is also possible to do the same for the
neutron-proton cross section. In order to describe the $T=1$ isospin channel,
it is necessary to consider also the exchange of the $\rho$ meson. This
calculation is presently under way, as well as the inclusion of the $\eta$ and
$\delta$ mesons of the Bonn OBE potential. 

Other interaction Lagrangians can be implemented using the same formalism,
in particular with (non linear) meson interactions or chiral symmetry.
A chiral symmetric interaction, although it is desirable from general 
considerations, has not yet reached the level of accuracy of the
popular one-boson exchange picture in the description of $N$-$N$
scattering, so that we did not consider it in this work. This field
however is in rapid evolution \cite{ORK94,FST97,SR97}. For example, 
the Nijmegen group \cite{SR97} quote a value of $\chi^2$ of 1.75 for 
a fit of the N-N scattering data with 12 free parameters. It will
bring a valuable piece of information to see how chirality modifies
the behavior of the cross section, in particular in the high density 
or high temperature range where the chiral restoration takes place.

\vskip 0.5cm

{\bf 6.3 Background velocity}

Also under consideration is the case when the background nuclear matter is
moving with a non-zero velocity with respect to the center of mass of the
collision. In this case, the behavior of the propagator in other regions of
the $\omega$-$q$ plane will be explored. For a given value of the energy of the
colliding particle and of the scattering angle, the values of the energy
$\omega$ and impulsion $q$ entering the expression of the polarizations is
given in section (3.2) as a function of the relative velocity.  In particular
(see Fig. 11), the region of the zero sound branches can be crossed. Such
branches arise from the mixed $\sigma$ - longitudinal $\omega$ part of the
dispersion relation \cite{DP91,DFH92}. If $(\omega,q)$ coincides
with a zero-sound solution $k^\mu_{zs}=(\omega_{zs},q_{zs})$ of 
the dispersion relation 
\beq
\left( k^2-\mu_\sigma^2+g_\sigma^2 \Pi_\sigma (k) \right) 
\left( k^2-\mu_\omega^2+g_\omega^2 \Pi_{\omega L} (k) \right) 
+ g_\sigma g_\omega \Pi_\times^2 (k) =0,
\eeq
there will be a critical value of the relative velocity $v_{crit}$ for which
the real part of the dispersion relation vanishes. For relative velocities 
near $v_{crit}$ the real part of the dispersion, which appears in the 
denominator of the propagators, is reduced. As a consequence, the
cross section will be enhanced.

A more detailed analysis of this and other effects of the relative velocity
of the center of mass of the collision with respect to the background
will be published separately.

\vskip 0.5cm

{\bf 6.4 $\Delta$ resonance}

An important extension of the model should be the inclusion of the $\Delta$
resonance. The $\Delta$ and higher mass resonances will appear at several
levels of the calculation: not only as the product of inelastic scattering
processes, but also in a virtual $\Delta$-hole pair which will modify the
propagation properties of the mesons and in particular of the pion. At high
densities and temperature, the nuclear matter turns into the so-called
resonance matter with a softer equation of state. 

At the relativistic energies we are considering, a large number 
of $\Delta(1232)$ are expected to be produced through the inelastic
processes $N + N \rightarrow \Delta +N$ and $N+N\rightarrow \Delta+\Delta$,
and reabsorbed by the inverse reactions. In the vacuum, the inelastic 
channel begins to open for laboratory energies $T_{lab} \geq$
350 MeV; at $T_{lab} \geq 800 MeV$, the elastic and inelastic 
contributions to the total nucleon-nucleon cross section are
roughly of the same size. The production {\it vs.} 
absorption rates
of $\Delta$ resonances is an important input in the calculation
of the rapidity distributions $d N_\pi/dy$ and $d N_K/dy$ of the 
pions and kaons since these particles are produced through $\Delta$
decays $N+N\rightarrow N+\Delta\rightarrow N+N+\pi$ and $N+N
\rightarrow N+\Delta\rightarrow N+\Lambda+K$. These distributions
are the quantities actually accessible by experiment and are 
used to characterize the thermodynamical state of the piece of excited
nuclear matter formed in a relativistic heavy ion collision.
Scatterings $N+\Delta\rightarrow N+\Delta$ will also contribute 
to the transport properties.
 
It is in principle straightforward to extend our formalism to calculate 
the $N+N \rightarrow N+\Delta$ cross section once the dressed
meson propagators are known. The $\Delta$ is coupled to the nucleon
through $\Delta$-N-$\pi$ and $\Delta$-N-$\rho$ vertices. The
scattering matrix is obtained from terms such as 
\beq 
Tr[G_N(1) \Gamma_{\pi N N} G_N(3) \Gamma_{\pi N N}] \  
Tr[G_N(2) \Gamma_{\pi N \Delta} G_\Delta(4) \Gamma_{\pi N \Delta}]  \ 
D_\pi(31) D_\pi^*(31)
\eeq
... {\it etc} 
and the $\Delta$ decay width may in a first approximation be taken into
account by integrating over 
a mass distribution function of the Breit-Wigner form. In this expression,
$G_N$ and $G_\Delta$ are the dressed, on-shell propagators of the nucleon 
and $\Delta$ respectively, and $D_\pi$ is the screened pion propagator. 
Similar calculations were performed already (see {\it e.g.} 
\cite{HA94} in the vacuum and \cite{BBKL88,MLZZ97} with dressed
pion propagators).
Nevertheless, before doing this, we think that it is profitable to 
perform a more thorough study of the consequences of screened meson
propagation on the nucleon-nucleon interaction on the example of elastic
scattering.

An other way the presence of the $\Delta$ can influence our calculations
is in the description of the background nuclear matter. The $\Delta$ can 
also appear in an internal line and contribute to the dressing of the 
mesons which mediate the interaction. It cannot be excluded that the 
corrections brought about by the creation of delta-hole loops in the 
propagators of the mesons might be sizable. But due to uncertainties 
and ambiguities in the description of the $\Delta$, a prediction at the
quantitative level seems premature.

The quantum hadrodynamics Lagrangian of Serot and Walecka can be extended in
order to include the $\Delta$ resonance. The resulting (nonrenormalizable)
Lagrangian (see {\it e.g.} \cite{WBB89}) introduces the $\Delta$ with the
Rarita-Schwinger spinor $\Psi_\Delta^\mu$. The $\Delta$ interacts through the
exchange of $\sigma$ and $\omega$ like the nucleons, moreover the $\Delta$ 
couples to the nucleons through $\Delta$-N-$\pi$ and $\Delta$-N-$\rho$ 
vertices. A first uncertainty comes from the fact that the coupling constants 
of the $\Delta$ to the $\sigma$, $\omega$ and $\rho$ are not very well known, 
and there is not enough data to constrain these parameters. A more serious 
difficulty arises from the Rarita-Schwinger formalism itself. As a matter 
of fact, the propagator $\Psi_\Delta^\mu \Psi_{\Delta\, \mu}$ contains not 
only spin 3/2 but also spin 1/2 parts. There has been a controversy whether 
or not one should project out the 1/2 part and how. When the $\Delta$ is 
interacting, the theory becomes plagued by so-called off-shell ambiguities, 
{\it i.e.} new free parameters may be introduced at the vertices. Studies 
on the influence on the $\Delta$ propagation noticed a strong dependence 
in these off-shell parameters \cite{HWB92}. 

There exists an abundant litterature on the modification of the $\pi$ 
dispersion relation by the $\Delta$ (see {\it e.g.} \cite{OTW82,HWB92,AH94} ). 
This topic is not yet completely settled, especially concerning the 
relativistic treatment as discussed above. We have seen that the pion 
did not play a very important role in the proton-proton cross section. 
As a matter of fact, we checked this statement by simply replacing the 
dressed pion propagator by the free one, and observed a modification 
amounting to 5\% at most. For the neutron-proton cross section nevertheless, 
the pion propagator will contribute a sizable part, as well as the $\rho$.

On the other hand, the modifications to the $\sigma$ and $\omega$ dispersion
relations should be investigated in the case of $pp$ scattering as well
as for $np$ scattering. Additional polarizations are given by terms
such as:
\beq
\int d^4p\ Tr[G_\Delta(p+k/2) \Gamma_{\Delta\Delta\sigma} G_\Delta(p-k/2)
\Gamma_{\Delta\Delta\sigma}]
\eeq
As far as we know, the modification of other mesons 
than the pion by delta-hole loops has not yet been considered in the 
litterature; this will need to be investigated in future work.

\vskip 0.5cm

{\bf 6.5 Form factors}

The phenomenological form factors account for the fact that the nucleons 
are not pointlike, but have a quark gluon substructure. As a consequence,
very high momentum transfer $q$ would probe very short ranges at which
our meson exchange picture fails. The cutoff suppresses these unphysical
contributions. The standard monopole form  chosen in this work (see
Eq. (\ref{eq:(3.8)})) fits well the known data on the form factor in the 
vacuum.

At finite density and temperature, we kept this form factor unchanged 
according to our level of approximation. In fact, the form factor is
expected to be modified in the medium: the nucleons ``swell'' 
(EMC effect) with increasing density. In a simple bag model picture, 
this is due to a reduced bag pressure. Higher order calculations of
the vertex function \cite{AS92-ST95}, where the nucleon size is 
generated by a surrounding meson cloud, yield the same behavior.
Higher momentum exchanges will thus be more strongly suppressed
at higher density.

\vskip 1cm

{\bf APPENDIX A: Analytical formulae for the polarisations}

Without loss of generality, the explicit form of the polarization terms can be
given in the particular frame where $ \displaystyle k \equiv (\omega, 0, 0,
q)$. In this frame the only non-vanishing components of the polarizations are
\cite{DP91}:
 
\beq
    & \Pi_{\sigma \omega}\ ^0 (k)\ ; \
    \Pi_{\sigma \omega}\ ^3 (k)\ ; \
    \Pi_\omega\ ^{00} (k)\ ;\
    \Pi_\omega\ ^{03} (k)\ ;\
     \Pi_\omega\ ^{33} (k)\ ; \nonumber \\
    & \Pi_\omega\ ^{11} (k)\
    = \ \Pi_\omega\ ^{22} (k)\ ; \
    \Pi_\sigma (k)\ ; \ 
    \Pi_\pi (k)
    \label{eq:(A.1)}
\eeq

The expressions of these components are directly obtained from

\beq
    \Pi_\sigma (k)\ =\ 8 \pi \int d^4p\
    \left[ (p-k/2) (p+k/2)\ +\ M^2 \right] .
    \Sigma (k,p)
    \label{eq:(A.2)}
\eeq
\beq
    \Pi_{\sigma \omega}\ ^\mu (k)\ =\ 16 \pi M
    \int d^4p\ p^\mu . \Sigma (k,p)
    \label{eq:(A.3)}
\eeq
\beq
    \Pi_\omega\ ^{\mu \nu} (k)\ =\ & 16 \pi
    \int d^4p\ \left[ p^\mu p^\nu\
    -\ k^\mu k^\nu/4\ -\ (1/2)\
    (p^2\ -\ k^2/4\ -\ M^2) g^{\mu \nu} \right] . \Sigma (k,p)
    \label{eq:(A.4)}
\eeq
\beq
    \Pi_\pi (k)\ =\ 8 \pi \int d^4p\
    \left[ (p-k/2) (p+k/2)\ -\ M^2 \right] .
    \Sigma (k,p)
    \label{eq:(A.5)}
\eeq

\noindent 

where

\beq
    \Sigma (k,p)\ = \left[ f (p+k/2)\
    -\ f (p-k/2)
     \right]/(k.p \ - \ i \epsilon)
    \label{eq:(A.6)}
\eeq

\noindent and the $ i \epsilon $ term comes from the boundary prescription
for the fermionic dressed propagator which enters the loop expressions of the
polarizations. It gives the imaginary parts of these polarizations through the
well know formula $1/(x-i \epsilon) = {\cal P}/x +i \pi \delta(x)$, where
$\cal P$ is the principal part. The fermion distribution function  $ f (p) $
in (\ref{eq:(A.6)}) is given by 

\beq
    f(p)\  =\ (2\pi)^{-3}\
    \delta(p^2\ -\ M^2).
    \left[\Omega\ ^+\ +\ \Omega\ ^-\
    -\ H(-p_0) \right]\
    \label{eq:(A.7)}
\eeq

the $ \Omega$'s are the Fermi factors defined as :

\beq
    \Omega ^\pm \ =\ H(\pm p_0).
    \{1\ +\ {\rm exp}(\beta [E(\vec p)\
    \mp\ \mu]) \}^{-1}
    \label{eq:(A.8)}
\eeq

and $H(x)$ is the Heaviside step function. $ \beta $ is the inverse temperature
and $ \mu $ is the auxiliar chemical potential, related to the true chemical
potential by $ \mu_{true} \ =\ \mu \ -\ g_{\omega}.\omega_{0} $. The
quasi-particles are on the mass-shell: 

\beq
    M\ =\ m \ - \  g_\sigma \sigma\;
    \qquad
    E(\vec p)\ =\ \sqrt{}\ ({\vec p}^2\ +\ M^2)
    \label{eq:(A.9)}
\eeq

\vskip 0.5cm

{\bf A.1 Real part of polarization tensors}

The matter contributions to the real part of the polarization tensors is
inmediatly obtained from Eqs. (\ref{eq:(A.2)}) to (\ref{eq:(A.5)}) by excluding
the last term in Eqs. (\ref{eq:(A.7)}). This last term corresponds to the
quasi-nucleon distribution in the Dirac sea (vacuum distribution) and leads to
divergences in the calculation of the real parts of the polarization tensors
which must be renormalized. For the real part of the scalar polarization $
\Pi_\sigma (k) , $ the renormalized vacuum contribution has been obtained as
usually, by regularizing the infinite vacuum polarization (which contains
divergences proportionnal to $k^{2}$, $\sigma$ and $\sigma^2$) and adding to
the Lagrangian the counterterms which allow to cancel such divergences. The
finite undefined constants in the renormalized expression are determined by
imposing a on-mass-shell criterion to the real renormalized dispersion relation
for the scalar mesons at vanishing density \cite{DP91}. The final form of
the vacuum finite ($v.f.$) part of the scalar polarization is 

\beq
   \Pi_{\sigma v f} (k)\
    & =\ (2k^2\ -\ 8m^2) \theta_\sigma\
      +\ 8(M^2\ -\ k^2/4)
      \theta (k^2, M^2) \nonumber \\
    & +\ 8 (m^2\ -\ \mu_\sigma\ ^2/4)
      (\mu_\sigma\ ^2\ -\ k^2) \theta_{\sigma k}\
      +\ (6M^2\ -\ k^2)\ {\rm ln} (M/m)^2 \nonumber \\
    & -\ 2g^2 \sigma^2\
      \left[ 6\ +\ \mu_\sigma\ ^2/m^2\
      +\ 16m^2 \theta_{\sigma m}\
      +\ 8 (m^2\ -\ \mu_\sigma\ ^2/4)
      m^2 \theta_{\sigma mm} \right] \nonumber \\
    & -\ (M^2\ -\ m^2)\
      \left[ 6\ -\ \mu_\sigma\ ^2/m^2\
      +\ 8 \theta_\sigma\
      +\ 8 (m^2\ -\ \mu_\sigma\ ^2/4)
      \theta_{\sigma m} \right]
    \label{eq:(A.10)}
\eeq

and the function $ \theta (k^2, M^2) $ is given by the
integral:

\beq
    \theta (k^2, M^2)\ =\ \theta (y)\
    =\ y \int^\infty_0 dx/
    \left[(x^2\ +\ y). \sqrt{}(x^2\ +\ 1) \right] 
    \label{eq:(A.11)}
\eeq

\noindent with :
\[
    y\ \equiv\ 1\ -\ k^2/(4M^2).
\]

In writing down Eq. (\ref{eq:(A.10)}), we have introduced the notations:

\beq
    \theta_\sigma\
    & =\ \theta(\mu_\sigma\ ^2, m^2), \nonumber \\
    \theta_{\sigma m}\
    & =\ \partial \theta (k^2, M^2)/\partial M^2 , \nonumber \\
    \theta_{\sigma k}\
    & =\ \partial \theta (k^2, M^2)/\partial k^2 \nonumber \\
    \theta_{\sigma mm}\
    & =\ \partial^2 \theta (k^2, M^2) / \partial (M^2)^2.
    \label{eq:(A.12)}
\eeq
 
All these derivatives are to be calculated at the point $ k^2 \ = \ 
\mu_\sigma^2, $ $ M \ = \ m $.

As it can easily be seen, there is no vacuum contribution
to the mixing polarization part $ \Pi_{\sigma \omega}\ ^\mu (k).$
 
The finite vacuum contributions to the vector polarization tensor $ \Pi_\omega\
^{\mu \nu}(k), $ is obtained from Eq.(\ref{eq:(A.4)}) using dimensional
regularization and compensating the divergent contributions through the
appropriate counterterms \cite{DP91}. The final expression is 

\beq
    & \Pi_{\omega v f}\ ^{\mu \nu} (k)
    \ = \ (4/3 \pi) \left[k^2 {\rm ln} (M/m)
    - 2M^2 + (2M^2 + k^2) \theta (k^2, M^2)
    +\Gamma \right] \Delta^{\mu \nu} (k)
    \label{eq:(A.13)}
\eeq

\noindent where the projector $ \Delta^{\mu \nu} (k) $ is defined by:

\beq
    \Delta^{\mu \nu}(k)\ =\
    g^{\mu \nu}\ -\ k^\mu k^\nu/k^2
    \label{eq:(A.14)}
\eeq

Here $ \theta (k^2, M^2) $ is the same function defined in Eq.
(\ref{eq:(A.11)}) and $ \Gamma $ is a finite constant which, determined through
the on-shell renormalization conditions, becomes: 

\beq
    \Gamma\ =\ 2m^2\ -\ (2m^2\ +\ \mu_\omega\ ^2)
    \theta (\mu_\omega\ ^2, m^2)
    \label{eq:(A.15)}
\eeq
 
The renormalized vacuum contributions to the real part of the pion polarization
obtained from the same on-shell prescription is \cite{DP91}

\beq
    \Pi_{\pi v f} (k)\
    & =\ 2k^2 \left[ \theta_\pi\
      -\ \theta (k^2, M^2)\
      +\ \mu_\pi\ ^2 \theta_{\pi k} \right]\
      -\ 2\mu_\pi\ ^4 \theta_{\pi k} \nonumber \\
    & +\ (2M^2\ -\ k^2)\ {\rm ln}\ (M/m)^2\
      -\ (2\ -\ \mu_\pi\ ^2/m^2\ -\ 2\mu_\pi\ ^2 \theta_{\pi m})
      \ (M^2\ -\ m^2) \nonumber \\
    & -\ 2g^2 \sigma^2\
      (2\ +\ \mu_\pi\ ^2/m^2\
      -\ 2\mu_\pi\ ^2 m^2 \theta_{\pi mm})
    \label{eq:(A.16)}
\eeq

where the definition of the derivatives of the functions $ \theta $ is the same
as in Eqs. (\ref{eq:(A.11)}) with the replacement of $ \mu_{\sigma} $ by $
\mu_{\pi} $. 

\vskip 0.5cm

{\bf A.2 Imaginary part of polarization tensors.}

The only non-vanishing components of the imaginary parts of the polarization
tensors are: 

\beq
   \Pi_{\pi,I} (k) \ & ; & \\
   \Pi_{\sigma,I} (k)\ & ; &  \\ 
   \Pi_{\sigma \omega,I} ^3 (k)\ & = & \ 
    (\omega/q) \Pi_{\sigma \omega,I}^0 (k)\ ;  \\
   \Pi_{\omega,I}^{33} (k)\ & = & \ (\omega/q) \Pi_{\omega,I}^{03} (k)\ = \
    (\omega^2/q^2) \Pi_{\omega,I}^{00} (k)\ ; \\
   \Pi_{\omega,I}^{11} (k)\
    & =& \ \Pi_{\omega,I}^{22} (k) 
    \label{eq:(A.17)}
\eeq

the relations between the vector-mixing and tensor components being a
consequence of the baryonic charge conservation. The expression for the finite
temperature imaginary part of the scalar polarization is: 

\beq
    \Pi_{\sigma,I} (\omega,q) \ = \ \Delta (\omega^2 \ - \ q^2)/q 
    \int_{M}^{\infty} dE \left[ H(A^+) \ - \ H(A^-) \right] 
    \left[ \Omega^+ \ - \ \Omega^- \ - \ 1 \right]
    \label{eq:(A.18)}
\eeq

where 

\beq
    A^{\pm} \ = \ q \sqrt{E^2 \ - \ M^2} \ - \ \mid \omega E \pm 
    (\omega^2 \ - \ q^2)/2 \mid
    \label{eq:(A.19)}
\eeq
\beq
    \Omega ^\pm \ =\   \{1\ +\ {\rm exp}(\beta [E \mp\ \mu]) \}^{-1}
    \label{eq:(A.20)}
\eeq
\beq
    \Delta \ = \ 1 \ - 4 M^2 /(\omega^2 \ - \ q^2) 
    \label{eq:(A.21)}
\eeq

The three terms inside brackets in Eq. (\ref{eq:(A.18)}) correspond
respectively to the contributions of baryons, antibaryons and vacuum. Here and
in the next formulae the vacuum contributions are finite. 

The imaginary part of the pion polarization is obtained from the imaginary part
of the $ \sigma $ polarization through 

\beq
    \Pi_{\pi,I} (\omega,q) \ =\ \Pi_{\sigma,I} (\omega,q) /\Delta
    \label{eq:(A.22)}
\eeq

The independent imaginary component of the mixing polarization vector is

\beq
     \Pi_{\sigma \omega,I}^0 (\omega,q) \ =\ {4M \over q} \   
    \int_{M}^{\infty} dE \ \left[ \left( E - {\omega\over 2}\right) 
    H(A^-) \ - \ \left( E + {\omega\over 2} \right) H(A^+) \right] \  
    (\Omega^+ \ + \ \Omega^- )
    \label{eq:(A.23)}
\eeq

where the vacuum contributions vanish.

The independent imaginary components of the $\omega$ polarization tensor are

\beq
    \Pi_{\omega,I}^{00} (\omega,q) & =  
    \displaystyle{ {4 \over q}\ \int_{M}^{\infty} } & dE \ 
    \left\{ \left[ \left( E - {\omega \over 2}\right) ^2  
    - {q^2 \over 4} \right] H(A^-) \ - \ \left[ \left( E + 
    {\omega\over 2}\right) ^2  - {q^2\over 4} \right] H(A^+) \right\}
    \nonumber \\
    & & \times \left( \Omega^+ \ - \ \Omega^- \ + \ 1 \right)
    \label{eq:(A.24)}
\eeq

and

\beq
    \Pi_{\omega,I}^{11} (\omega,q) & =  
    \displaystyle{ {2\over q} \ \int_{M}^{\infty} } & dE \   
    \left\{ \left[ \left( E\ - \ {\omega\over2} \right) \left( 1 \ - \ 
    {\omega^2 \over q^2}\right) \ - \ M^2 \right] H(A^-) \right. \nonumber \\ 
    & & \left. - \ \left[ \left( E + {\omega\over 2}\right) \left( 1 \ - \ 
    {\omega^2\over q^2}\right) \ - \ M^2 \right] H(A^+) \right\} \ 
    \left( \Omega^+ - \Omega^- + 1 \right)
    \label{eq:(A.25)}
\eeq

\vskip 1cm

{\bf APPENDIX B: Analytical expression of the cross section}

The expression of the spin-averaged nucleon-nucleon cross section 
results from taking the square of the transition matrix Eq. (\ref{eq:(3.10)})
and using the relation $\sum_{s} u(p,s) \overline u(p,s)=\gamma.p+m$. We use
the short notation $S_i=\gamma.p_{i\, *}+m_*$.

The calculation of the transition matrix is straightforward.
We give here the result as a function of the Mandelstam variables
and of the relative background velocity. Additional terms
appear due to $\sigma$-$\omega$ mixing, for example for the
direct term:

\beq
\overline{|{\cal M}|^2} & \ni \displaystyle{g_\sigma^2 g_\omega^2 \over 4} 
     &  \left\{ {\cal F}_\sigma^2(k_{31\, *})\ {\cal F}_\omega^2(k_{31\, *})\ 
     \left( \Tr[S_3 \gamma_\mu S_1 \gamma_\alpha]\ \Tr[S_4 S_2]
     \right. \right. \nonumber \\
& &  \left. \left. \phantom{ghost} 
     + \Tr[S_3 \gamma_\mu S_1]\ \Tr[S_4 S_2 \gamma_\alpha] 
     + \Tr[ S_3 S_1 \gamma_\alpha]\ \Tr[S_4 \gamma_\mu S_2] 
     \right. \right. \nonumber \\
& &  \left. \left. \phantom{ghost}
     + \Tr[ S_3 S_1]\ \Tr[S_4 \gamma_\mu S_2 \gamma_\alpha] \right)
     G_{\sigma\omega}^\mu(k_{31\, *})\ G_{\sigma\omega}^\alpha(k_{31\, *})^*
     \right. \nonumber \\
&  + \displaystyle{g_\sigma^3 g_\omega\over 4} & 
    \left\{ {\cal F}_\sigma^3(k_{31\, *})\ 
    {\cal F}_\omega^2(k_{31\, *}) \left( \Tr[S_3 S_1 \gamma_\alpha]\ 
    \Tr[S_4 S_2] \right. \right. \nonumber \\
& & \left. \left. \phantom{ghost}
    + \Tr[S_3 S_1]\ \Tr[ S_4 S_2 \gamma_\alpha] \right) G_\sigma(k_{31\, *})
    \ G_{\sigma\omega}^\alpha(k_{31\, *})^* \right. \nonumber \\
& & \left. + {\cal F}_\sigma^3(k_{31\, *})\ {\cal F}_\omega^2(k_{31\, *}) 
    \left( \Tr[S_3 \gamma_\alpha S_1]\ \Tr[ S_4 S_2] 
    \right. \right. \nonumber \\
& & \left. \left. \phantom{ghost}
    + \Tr[ S_3 S_1]\ \Tr[ S_4 S_2 \gamma_\alpha] \right) 
    G_\sigma(k_{31\, *})^* \ G_{\sigma\omega}^\alpha(k_{31\, *}) 
    \right\} \nonumber \\
&  + \displaystyle{g_\sigma g_\omega^3\over 4} & 
    \left\{ {\cal F}_\sigma(k_{31\, *})\ {\cal F}_\omega^3(k_{31\, *})
    \left( \Tr[S_3 \gamma_\mu S_1 \gamma_\alpha]\ \Tr[ S_4 \gamma_\nu S_2]
    \right. \right. \nonumber \\
& & \left. \left. \phantom{ghost}
    +\Tr[S_3 \gamma_\mu S_1]\ \Tr[S_4 \gamma_\nu S_2 \gamma_\alpha] \right) 
    G_\omega^{\mu\nu}(k_{31\, *})\ G_{\sigma\omega}^\alpha(k_{31\, *})^*
    \right. \nonumber \\
& &  + \left. {\cal F}_\sigma(k_{31\, *})\ {\cal F}_\omega^3(k_{31\, *})
    \left( \Tr[S_3 \gamma_\alpha S_1 \gamma_nu]\ \Tr[S_4 S_2 \gamma_\mu]
    \right. \right. \nonumber \\
& & \left. \left. \phantom{ghost}
    + \Tr[S_3 S_1 \gamma_\nu] \Tr[S_4 \gamma_\alpha S_2 \gamma_\mu] \right)
   G_\omega^{\mu\nu}(k_{31\, *})^*\ G_{\sigma\omega}^\alpha(k_{31\, *})
   \right\} \nonumber \\
& + \displaystyle{g_\sigma g_\omega g_\pi^2 \over 4} & 
\left\{ {\cal F}_\sigma(k_{31\, *})\ {\cal F}_\omega(k_{31\, *}) \ 
    {\cal F}_\pi^2(k_{31\, *}) \left( \Tr[ S_3  \gamma_\mu S_1 \gamma_5]\ 
    \Tr[S_4 S_2 \gamma_5] \right. \right. \nonumber \\
& & \left. \left. \phantom{ghost} 
    + \Tr[ S_3 S_1 \gamma_5]\ \Tr[  S_4 \gamma_\mu S_2 \gamma_5] \right)
    G_{\sigma\omega}^{\mu}(k_{31\, *})\ G_\pi(k_{31\, *})^* 
    \right. \nonumber \\
& & \left. + {\cal F}_\sigma(k_{31\, *})\ {\cal F}_\omega(k_{31\, *}) \
    {\cal F}_\pi^2(k_{31\, *}) \left( \Tr[ S_3 \gamma_5  S_1 \gamma_\mu]\
    \Tr[S_4 \gamma_5 S_2] \right. \right. \nonumber \\
& & \left. \left. \phantom{ghost}
    +\Tr[ S_3 \gamma_5 S_1]\ \Tr[ S_4 \gamma_5 S_2 \gamma_\mu] \right)
    G_{\sigma\omega}^{\mu}(k_{31\, *})^*\ G_\pi(k_{31\, *}) 
    \right\} \nonumber 
\eeq

In this formula, the ${\cal F}(k)$ are the form factors given by Eq.
(\ref{eq:(3.8)}) and $k_{31}$ is the quadrivector for momentum transfer
in the direct diagram $k_{31\, *}^\mu=p_{1\, *}^\mu-p_{3\, *}^\mu$

The Mandelstam variables are related to the incident $p_{1\, *}$, $p_{2\, *}$
and outcoming $p_{3\, *}$, $p_{4\, *}$ impulsions by
\beq
s &=& (p_{1\, *}+p_{2\, *})^2 \\
t &=& (p_{1\, *}-p_{3\, *})^2 \\
u &=& (p_{1\, *}-p_{4\, *})^2 
\eeq
with $s+t+u=4\ m_*^2$.
 
When the incident and outcoming particles are on their effective mass-shell 
$p_{i\, *}^2=m_*^2$, and if take into account the energy momentum conservation 
$p_{1\, *}^\mu+p_{2\, *}^\mu = p_{3\, *}^\mu+p_{4\, *}^\mu$, we have
\beq
s &=& 2\ (m_*^2+p_{1\, *}.p_{2\, *}) = 4\ E^2 \\
t &=& 2\ (m_*^2-p_{1\, *}.p_{3\, *}) = 2\ (m_*^2-E^2)(1-\cos\theta) \\
u &=& 2\ (m_*^2-p_{1\, *}.p_{4\, *}) = 2\ (m_*^2-E^2)(1+\cos\theta) 
\eeq

If the relative velocity between the background nuclear matter and the 
center of mass of the collision $u^\mu=(\gamma ,\gamma v \sin 
\alpha\cos\varphi,\gamma v \sin\alpha\cos\varphi,\gamma v\cos \alpha)$ 
is finite, and when calculating terms containing the longitudinal part 
of the vector meson propagator, we will also have components along the 
vectors
\beq
\eta_\ominus^\mu=u^\mu-{\eta.u\over q_{31}^2} q_{31}^\mu \qquad {\rm and}
\qquad \eta_\oplus^\mu=u^\mu-{\eta.u\over q_{41}^2} q_{41}^\mu
\nonumber
\eeq
We note the projections of the particle momenta on these
directions $p_1^\ominus=p_1^\mu.\eta_{\ominus\mu}$, {\it etc.} 
They are related to the Mandelstam variables defined above
and to the components of the relative velocity through
\beq
p_1^\ominus &=p_3^\ominus =& \sqrt{t\over t-\omega_\ominus^2} \left[ 
      \gamma \sqrt{s\over 4} -\gamma v \cos\alpha\sqrt{{s\over 4}-m_*^2} 
      - {\omega_\ominus \over 2} \right] \nonumber \\
p_2^\ominus &=p_4^\ominus =& \sqrt{t\over t-\omega_\ominus^2} \left[ 
      \gamma \sqrt{s\over 4} +\gamma v \cos\alpha\sqrt{{s\over 4}-m_*^2} 
      + {\omega_\ominus \over 2} \right] \nonumber \\
p_1^\oplus &=p_4^\oplus =& \sqrt{u\over u-\omega_\oplus^2} \left[ 
      \gamma \sqrt{s\over 4} -\gamma v \cos\alpha\sqrt{{s\over 4}-m_*^2} 
      - {\omega_\oplus \over 2} \right] \nonumber \\
p_2^\oplus &=p_3^\oplus =& \sqrt{u\over u-\omega_\oplus^2} \left[ 
      \gamma \sqrt{s\over 4} +\gamma v \cos\alpha\sqrt{{s\over 4}-m_*^2} 
      + {\omega_\oplus \over 2} \right] \nonumber 
\eeq
The $\omega_\ominus$ and $\omega_\oplus$ were defined in section
(3.2). In terms of the Mandelstam variables,
\beq
\omega_\ominus &=& \gamma v \sqrt{{s\over 4}-m_*^2} 
\left[ \sin\alpha \sin\varphi \sqrt{1-\left( \displaystyle{t-u\over s}
\right) } -\cos\alpha(1-{t-u\over s}) \right] \nonumber \\
\omega_\oplus &=& \gamma v \sqrt{{s\over 4}-m_*^2} 
\left[ - \sin\alpha \sin\varphi \sqrt{1-\left( \displaystyle{t-u\over s}
\right) } -\cos\alpha(1+{t-u\over s}) \right]\nonumber
\eeq

We split for convenience here the propagator of the $\omega$ vector
mesons in a transverse part $G_{\omega T}$ and a part along the vector
$\eta$:
 $G_{\omega N}(t)=G_{\omega L}(t)-G_{\omega T}(t)$.

The spin-averaged scattering matrix for the Coulomb substracted 
elastic nucleon-nucleon scattering is given by:

\beq
{d \sigma_{pp} \over d \Omega} &=& {d \sigma_{nn} \over d \Omega} \\
    &=& {1 \over 64 \pi^2 s} \left[ {\cal S}_{4\sigma} + {\cal S}_{4\omega}
    + {\cal S}_{2\sigma 2\omega} + {\cal S}_{1\sigma 3\omega} 
    + {\cal S}_{3\sigma 1\omega} + {\cal S}_{4\pi} + {\cal S}_{2\pi 2\sigma}
    + {\cal S}_{2\pi 2\omega} + {\cal S}_{2\pi 1\sigma 1\omega} \right]
    \nonumber
\eeq
with ($e$, $d$ and $i$ are the direct, exchange and interference terms)
\beq
{\cal S}_{4\sigma} & = & {\cal S}_{4\sigma}^{d} + {\cal S}_{4\sigma}^{e} 
      + {\cal S}_{4\sigma}^{i} \nonumber \\
{\cal S}_{4\sigma}^{d} & = & g_\sigma^4\ {\cal F}^4_\sigma(t)\ 
      \left[\, (4\ m_*^2-t)^2\right] \ G_\sigma(t)\ G^*_\sigma(t) \nonumber \\
{\cal S}_{4\sigma}^{e} & = & g_\sigma^4\ {\cal F}^4_\sigma(u)\ 
      \left[\, (4\ m_*^2-u)^2\right] \ G_\sigma(u)\ G^*_\sigma(u) \nonumber \\
{\cal S}_{4\sigma}^{i} & = & - g_\sigma^4\ {\cal F}^2_\sigma(t)\ 
      {\cal F}^2_\sigma(u) \left[ {t^2+u^2-s^2 \over 4} + 4\ m_*^2(s-\ m_*^2)
      \right] . \left\{ G_\sigma(t)\ G^*_\sigma(u) + G^*_\sigma(t)\ 
      G_\sigma(u) \right\} \nonumber \\
& & \nonumber \\
& & \nonumber \\
{\cal S}_{4\omega} & = & {\cal S}_{4\omega}^{d} + {\cal S}_{4\omega}^{e} 
                      + {\cal S}_{4\omega}^{i} \nonumber \\
& & \nonumber \\
{\cal S}_{4\omega}^{d} & = & g_\omega^4 {\cal F}_\omega(t)^4 
\left( 4 \left[ (s-u)^2+t(t+8 m_*^2) \right].
                   \left\{ G_{\omega T}(t) G^*_{\omega T}(t) \right\}
                  \right. \nonumber \\
& & \phantom{g_\omega^4 {\cal F}_\omega(t)^4}
      + 4 \left[ 4 t (p_{1\ominus}^2 + p_{4\ominus}^2)
            +16 p_{1\ominus}^2 p_{4\ominus}^2 +t^2 \right] 
            \left\{ G_{\omega N}(t) G_{\omega N}^*(t) \right\} 
       \nonumber \\
& & \phantom{g_\omega^4 {\cal F}_\omega(t)^4} 
      +4 \left[ 4 (s-u) p_{1\ominus} p_{4\ominus}
      +4 t( p_{1\ominus}^2+p_{4\ominus}^2) + t^2 \right] \times
      \nonumber \\
& & \left. \phantom{g_\omega^4 {\cal F}_\omega(t)^4} \phantom{ghost}
     \times \left\{ G_{\omega N}(t) G_{\omega T}^*(t) +G_{\omega N}^*(t) 
      G_{\omega T} (t)\right\} \right) \nonumber \\
{\cal S}_{4\omega}^{e} & = & g_\omega^4 {\cal F}_\omega(u)^4 
\left( 4 \left[ (s-t)^2+u(u+8 m_*^2) \right] .
                   \left\{ G_{\omega T}(u) G_{\omega T}^*(u) \right\}
                  \right. \nonumber \\
& & \phantom{g_\omega^4 {\cal F}^4_\omega(u)}
    + 4 \left[ 4 u (p_{1\oplus}^2 + p_{4\oplus}^2)
            +16 p_{1\oplus}^2 p_{4\oplus}^2 +u^2 \right] 
            \left\{ G_{\omega N}(u) G_{\omega N}^*(u) \right\}
       \nonumber \\
& & \phantom{g_\omega^4 {\cal F}^4_\omega(u)}
     +4 \left[ 4 (s-t) p_{1\oplus} p_{4\oplus}
      +4 u( p_{1\oplus}^2+p_{4\oplus}^2) + u^2 \right] \times
      \nonumber \\
& &  \left. \phantom{g_\omega^4 {\cal F}^4_\omega(u)} \phantom{ghost}
      \times \left\{ G_{\omega N}(u) G_{\omega T}^*(u) +G_{\omega N}^*(u) 
      G_{\omega T} (u) \right\} \right) \nonumber \\
{\cal S}_{4\omega}^{i} & = & g_\omega^4 {\cal F}^2_\omega(u) 
     {\cal F}^2_\omega(t) \left(  - \left[ 8 (s-2 m_*^2)(6 m_*^2-s)\right]. 
      \left\{ G_{\omega T}(t) G_{\omega T}^*(u)
       +G_{\omega T}^*(t) G_{\omega T}(u) \right\} \right. \nonumber \\
& &  \phantom{g_\omega^4 {\cal F}^2_\omega(u)}
     -\left[ 8 (4 m_*^2 -s) (p_{1\ominus}p_{4\ominus}+p_{1\oplus}p_{3\oplus})
     -4 t (p_{1\ominus}^2+p_{4\ominus}^2) - 4 u (p_{1\oplus}^2+
     p_{3\oplus}^2) \right. \nonumber \\
& & \left. \phantom{g_\omega^4 {\cal F}^2_\omega(u)}
     +32 p_{1\ominus} p_{4\ominus} p_{1\oplus} p_{3\oplus}
     - 2 u t (2 \eta_\ominus.\eta_\oplus -1) \right]
     \times \nonumber \\
& & \phantom{g_\omega^4 {\cal F}^2_\omega(u)} \phantom{ghost} 
     \times \left\{ G_{\omega N}(t) G_{\omega N}^*(u) + G_{\omega N}^*(t) 
     G_{\omega N}(u)\right\} \nonumber \\
& & \phantom{g_\omega^4 {\cal F}^2_\omega(u)}
      - \left[ 4 s (p_{1\oplus}-p_{3\oplus})^2 -8 u (p_{1\oplus}^2 +
        p_{3\oplus}^2) +4 t (p_{1\oplus}+p_{3\oplus})^2 -8 u m_*^2 
       \right. \nonumber \\
& & \phantom{g_\omega^4 {\cal F}^2_\omega(u)}
       -16 m_*^2 (p_{1\oplus}^2+p_{3\oplus}^2-4 p_{1\oplus} p_{3\oplus})
       -2(s-2 m_*^2)(s-u)+2(2 m_*^2-t)(4 m_*^2-t) 
       \nonumber \\
& & \left. \phantom{g_\omega^4 {\cal F}^2_\omega(u)} 
       +2 u t(1+\displaystyle{2\over t}
        (p_{1\oplus}-p_{3\oplus})^2) \right] 
\left\{ G_{\omega N}(u) 
        G_{\omega T}^*(t) + G_{\omega N}^*(u) G_{\omega T}(t) \right\}
       \nonumber \\
& &  \phantom{g_\omega^4 {\cal F}^2_\omega(u)}
  - \left[ 4 s (p_{1\ominus}-p_{4\ominus})^2 -8 t (p_{1\ominus}^2 +
        p_{4\ominus}^2) +4 u (p_{1\ominus}+p_{4\ominus})^2 -8 t m_*^2
        \right. \nonumber \\
& & \phantom{g_\omega^4 {\cal F}^2_\omega(u)}
       -16 m_*^2 (p_{1\ominus}^2+p_{4\ominus}^2-4 p_{1\ominus} p_{4\ominus})
       -2(s-2 m_*^2)(s-t)+2(2 m_*^2-u)(4 m_*^2-u) 
        \nonumber \\
& &  \left. \left.  \phantom{g_\omega^4 {\cal F}^2_\omega(u)}
       +2 u t (1+\displaystyle{2\over u}
        (p_{1\ominus}-p_{4\ominus})^2) \right] 
     \left\{ G_{\omega N}(t) G_{\omega T}^*(u) + G_{\omega N}^*(t) 
     G_{\omega T}(u) \right\} \right) \nonumber \\
& & \nonumber \\
& & \nonumber \\
{\cal S}_{2\sigma 2\omega} & = & {\cal S}_{2\sigma 2\omega}^{d} + 
        {\cal S}_{2\sigma 2\omega}^{e} + {\cal S}_{2\sigma 2\omega}^{i} 
        \nonumber \\
& & \nonumber \\
{\cal S}_{2\sigma 2\omega}^{d} & = & g_\sigma^2 g_\omega^2 
      {\cal F}^2_{\sigma}(t) {\cal F}^2_{\omega}(t) \left( [-16 m_*^2 (s-u)].
      \left\{ G_\sigma(t) G_{\omega T}^*(t)+G_\sigma^*(t) G_{\omega T}(t) 
      \right\}
      \right. \nonumber \\
& &             \phantom{g_\sigma^2 g_\omega^2 {\cal F}^2_{\sigma}(t) }
        +\left[ -64 m_*^2 p_{1\ominus} p_{4\ominus} \right] .
        \left\{ G_\sigma(t) G_{\omega N}^*(t)+G_\sigma^*(t) G_{\omega N}(t) 
         \right\} \nonumber \\
& &            \left. \phantom{g_\sigma^2 g_\omega^2 {\cal F}^2_{\sigma}(t)}
        +\left[ 16(4 m_*^2-t)(\displaystyle{t\over 2}+p_{1\ominus}^2+
         p_{4\ominus}^2) +128 m_*^2 p_{1\ominus}p_{4\ominus} \right] .
        \left\{ G_\times(t) G_\times^*(t) \right\}
                \right) \nonumber \\
{\cal S}_{2\sigma 2\omega}^{e} & = & g_\sigma^2 g_\omega^2 
      {\cal F}^2_{\sigma}(u) {\cal F}^2_{\omega}(u) \left( [-16 m_*^2 (s-t)].
      \left\{ G_\sigma(u) G_{\omega T}^*(u)+G_\sigma^*(u) G_{\omega T}(u) 
      \right\} \right. \nonumber \\
& &            \phantom{g_\sigma^2 g_\omega^2 {\cal F}^2_{\sigma}(u) }
        +[-64 m_*^2 p_{1\oplus} p_{3\oplus}].
        \left\{ G_\sigma(u) G_{\omega N}^*(u)+G_\sigma^*(u) G_{\omega N}(u) 
         \right\} \nonumber \\
& &            \left. \phantom{g_\sigma^2 g_\omega^2 {\cal F}^2_{\sigma}(u) }
        +\left[ 16(4 m_*^2-u)(\displaystyle{u\over 2}+p_{1\oplus}^2+
         p_{3\oplus}^2) +128 m_*^2 p_{1\oplus}p_{3\oplus} \right] .
        \left\{ G_\times(u) G_\times^*(u) \right\}
\right) \nonumber \\
{\cal S}_{2\sigma 2\omega}^{i} & = & g_\sigma^2 g_\omega^2 
          {\cal F}^2_{\sigma}(t) {\cal F}^2_{\omega}(u) \left(
          \left[ 4(24 m_*^2-4 m_*^2 s -10 m_*^2 t+t^2 ) \right] .
        \left\{ G_\sigma(t) G_{\omega T}^*(u)+G_\sigma^*(t) G_{\omega T}(u) 
        \right\} \right. \nonumber \\
& &  \phantom{g_\sigma^2 g_\omega^2 {\cal F}^2_{\sigma}(t) }
          +\left[ 2u(s+u)+4 u(p_{1\oplus}^2+p_{3\oplus}^2)+8 s p_{1\oplus} 
           p_{3\oplus} \right] \times \nonumber \\
& & \left. \phantom{g_\sigma^2 g_\omega^2 {\cal F}^2_{\sigma}(t)} 
        \phantom{fat ghost}  \times
        \left\{ G_\sigma(t) G_{\omega N}^*(u)+G_\sigma^*(t) G_{\omega N}(u) 
        \right\} \right) \nonumber \\
& & +g_\sigma^2 g_\omega^2 
          {\cal F}^2_{\sigma}(u) {\cal F}^2_{\omega}(t) \left(
          \left[ 4(24 m_*^2-4 m_*^2 s -10 m_*^2 u+u^2 ) \right] .
        \left\{ G_\sigma(u) G_{\omega T}^*(t)+G_\sigma^*(u) G_{\omega T}(t) 
        \right\} \right. \nonumber \\
& &  \phantom{g_\sigma^2 g_\omega^2 {\cal F}^2_{\sigma}(u) }
          +\left[ 2 t(s+t)+4 t(p_{1\ominus}^2+p_{4\ominus}^2)+8 s p_{1\ominus} 
           p_{4\ominus} \right] \times \nonumber \\
& & \left. \phantom{g_\sigma^2 g_\omega^2 {\cal F}^2_{\sigma}(u)} 
        \phantom{fat ghost}  \times
        \left\{ G_\sigma(u) G_{\omega N}^*(t)+G_\sigma^*(u) G_{\omega N}(t) 
        \right\} \right) \nonumber \\
& & -g_\sigma^2 g_\omega^2 {\cal F}_{\sigma}(t){\cal F}_{\sigma}(u)
           {\cal F}_{\omega}(u){\cal F}_{\omega}(t)
       \left[ 32 m_*^2(p_{1\ominus}+p_{4\ominus})(p_{1\oplus}+p_{3\oplus})
        +8 u t\ \eta_\ominus.\eta_\oplus \right] \times \nonumber \\
& & \phantom{-g_\sigma^2 g_\omega^2 {\cal F}^2_{\sigma}u} 
           \phantom{fat ghost}
           \times \left\{ G_\times(u) G_\times^*(t)+G_\times^*(u) G_\times(t)
           \right\} \nonumber \\ 
& & \nonumber \\
& & \nonumber \\
{\cal S}_{1\sigma 3\omega} & = & {\cal S}_{1\sigma 3\omega}^{d} +
        {\cal S}_{1\sigma 3\omega}^{e} + {\cal S}_{1\sigma 3\omega}^{i}
        \nonumber \\
& & \nonumber \\
{\cal S}_{1\sigma 3\omega}^{d} & = & g_\sigma g_\omega^3
             {\cal F}_\sigma(t) {\cal F}_\omega^3(t) \left(
          \left[ 32 m_*\left( (s-u)p_{1\ominus}+t p_{4\ominus} \right) \right] 
          . \left\{ G_\times(t) G_{\omega T}^*(t) +G_\times^*(t) 
          G_{\omega T}(t) \right\} \right. \nonumber \\
& &  \left. \phantom{g_\sigma g_\omega^3 {\cal F}_\sigma(t) } 
       + \left[ 16 m_* (p_{1\ominus}+p_{4\ominus})
       (t+4p_{1\ominus} p_{4\ominus}) \right].
       \left\{ G_\times(t) G_{\omega N}^*(t) +G_\times^*(t)
          G_{\omega N}(t) \right\} \right) \nonumber \\
{\cal S}_{1\sigma 3\omega}^{e} & = & g_\sigma g_\omega^3
             {\cal F}_\sigma(u) {\cal F}_\omega^3(u) \left(
          \left[ 32 m_*\left( (s-t)p_{1\oplus}+u p_{3\oplus} \right) \right] 
          . \left\{ G_\times(u) G_{\omega T}^*(u) +G_\times^*(u) 
          G_{\omega T}(u) \right\} \right. \nonumber \\
& &  \left. \phantom{g_\sigma g_\omega^3 {\cal F}_\sigma(u) } 
       + \left[ 16 m_* (p_{1\oplus}+p_{3\oplus})
       (u+4 p_{1\oplus} p_{3\oplus}) \right].
       \left\{ G_\times(u) G_{\omega N}^*(u) +G_\times^*(u)
          G_{\omega N}(u) \right\} \right) \nonumber \\
{\cal S}_{1\sigma 3\omega}^{i} & = & -g_\sigma g_\omega^3 
          {\cal F}_\sigma(t) {\cal F}_\omega(t){\cal F}^2_\omega(u) 
          \left( \left[ 16 m_*(p_{1\ominus}+p_{4\ominus})(u-t+2 m_*^2)
          \right]. \left\{ G_\times(t) G_{\omega T}^*(u) +G_\times^*(t)
          G_{\omega T}(u) \right\} \right. \nonumber \\
& & \left. \phantom{g_\sigma g_\omega^3 {\cal F}_\sigma(t) } 
          + \left[ 8 m_* (p_{1\ominus}+p_{4\ominus})(4 p_{1\oplus}p_{3\oplus}
          +u) \right] .\left\{ G_\times(t) G_{\omega N}^*(u) +G_\times^*(t)
          G_{\omega N}(u) \right\} \right) \nonumber \\
& & -g_\sigma g_\omega^3 
          {\cal F}_\sigma(u) {\cal F}_\omega(u){\cal F}^2_\omega(t) 
          \left( \left[ 16 m_*(p_{1\oplus}+p_{3\oplus})(t-u+2 m_*^2)
          \right]. \left\{ G_\times(u) G_{\omega T}^*(t) +G_\times^*(u)
          G_{\omega T}(t) \right\} \right. \nonumber \\
& & \left. \phantom{g_\sigma g_\omega^3 {\cal F}_\sigma(t) } 
          + \left[ 8 m_* (p_{1\oplus}+p_{3\oplus})(4 p_{1\ominus}p_{4\ominus}
          +t) \right] .\left\{ G_\times(u) G_{\omega N}^*(t) +G_\times^*(u)
          G_{\omega N}(t) \right\} \right) \nonumber \\
& & \nonumber \\
& & \nonumber \\
{\cal S}_{3\sigma 1\omega} & = & {\cal S}_{3\sigma 1\omega}^{d} +
        {\cal S}_{3\sigma 1\omega}^{e} + {\cal S}_{3\sigma 1\omega}^{i}
        \nonumber \\
& & \nonumber \\
{\cal S}_{3\sigma 1\omega}^{d} & = & -g_\sigma^3 g_\omega {\cal F}^3_\sigma(t)
        {\cal F}_\omega(t) \left[ 16 m_*(4m_*^2-t)(p_{1\ominus}+
        p_{4\ominus}) \right] .\left\{ G_\times(t) G_\sigma^*(t) +G_\times^*(t)
          G_\sigma(t) \right\} \nonumber \\
{\cal S}_{3\sigma 1\omega}^{e} & = & -g_\sigma^3 g_\omega {\cal F}^3_\sigma(u)
        {\cal F}_\omega(u) \left[ 16 m_*(4m_*^2-u)(p_{1\oplus}+
        p_{3\oplus}) \right] .\left\{ G_\times(u) G_\sigma^*(u) +G_\times^*(u)
          G_\sigma(u) \right\} \nonumber \\
{\cal S}_{3\sigma 1\omega}^{i} & = & g_\sigma^3 g_\omega \left(
        {\cal F}^2_\sigma(t){\cal F}_\sigma(u) {\cal F}_\omega(u) 
        \left[ 8 m_*(4 m_*^2-t)(p_{1\oplus}+p_{3\oplus}) \right] .
        \left\{ G_\sigma(t)G_\times^*(u)+G_\sigma^*(t)G_\times(u) \right\}
        \right. \nonumber \\
& & \left.  \phantom{g_\sigma^3 g_\omega} +{\cal F}^2_\sigma(u)
        {\cal F}_\sigma(t) {\cal F}_\omega(t)
        \left[ 8 m_*(4 m_*^2-u)(p_{1\ominus}+p_{4\ominus}) \right] .
        \left\{ G_\sigma(u)G_\times^*(t)+G_\sigma^*(u)G_\times(t) \right\}
        \right) \nonumber \\
& & \nonumber \\
& & \nonumber \\
{\cal S}_{4\pi} & = & {\cal S}_{4\pi}^{d} + {\cal S}_{4\pi}^{e} 
      + {\cal S}_{4\pi}^{i} \nonumber \\
& & \nonumber \\
{\cal S}_{4\pi}^{d} & = & g_\pi^4\ {\cal F}^4_\pi(t)\ 
      [\, t^2\, ]\ G_\pi(t)\ G_\pi(t)^* \nonumber \\
{\cal S}_{4\pi}^{e} & = & g_\pi^4\ {\cal F}^4_\pi(u)\ 
      [\, u^2\, ]\ G_\pi(u)\ G_\pi(u)^* \nonumber \\
{\cal S}_{4\pi}^{i} & = & g_\pi^4\ {\cal F}^2_\pi(t)\ 
      {\cal F}^2_\pi(u)\  [\, {ut\over 2}\, ]. \left\{ G_\pi(t)\ G_\pi(u)^* +
       G_\pi(t)^*\ G_\pi(u) \right\}
      \nonumber \\
& & \nonumber \\
& & \nonumber \\
{\cal S}_{2\pi 2\sigma} & = & {\cal S}_{2\pi 2\sigma}^{d} +
        {\cal S}_{2\pi 2\sigma}^{e} + {\cal S}_{2\pi 2\sigma}^{i}
        \nonumber \\
& & \nonumber \\
{\cal S}_{2\pi 2\sigma}^{d} & = & 0 \nonumber \\
{\cal S}_{2\pi 2\sigma}^{e} & = & 0 \nonumber \\
{\cal S}_{2\pi 2\sigma}^{i} & = & g_\pi^2\ g_\sigma^2 \left( 
      {\cal F}^2_\sigma(t)\ {\cal F}^2_\pi(u)\ [\, {(4\ m_*^2-t)u\over 2}\, ].
     \left\{ G_\sigma(t)\ G_\pi(u)^* 
      + G_\sigma(t)^*\ G_\pi(u) \right\} \right. 
      \nonumber \\
& &   \left. \phantom{g_\pi^2\ g_\sigma^2 } 
    + {\cal F}^2_\sigma(u)\ {\cal F}^2_\pi(t)\ [\, {(4\ m_*^2-u)t\over 2}\, ].
     \left\{ G_\sigma(u)\ G_\pi(t)^*
     + G_\sigma(u)^*\ G_\pi(t) \right\} \right)
\nonumber \\
& & \nonumber \\
& & \nonumber \\
{\cal S}_{2\pi 2\omega} & = & {\cal S}_{2\pi 2\omega}^{d} +
        {\cal S}_{2\pi 2\omega}^{e} + {\cal S}_{2\pi 2\omega}^{i}
        \nonumber \\
& & \nonumber \\
{\cal S}_{2\pi 2\omega}^{d} & = & 0 \nonumber \\
{\cal S}_{2\pi 2\omega}^{e} & = & 0 \nonumber \\
{\cal S}_{2\pi 2\omega}^{i} & = & -g_\omega^2 g_\pi^2 
          {\cal F}^2_\omega(t) {\cal F}^2_\pi(u) \left(
          \left[ 4 u(2 m_*^2-u) \right] .
          \left\{ G_{\omega T}(t)G_\pi^*(u)+G_{\omega T}^*(t)G_\pi(u) \right\}
          \right. \nonumber \\
& & \phantom{-g_\omega^2 g_\pi^2 {\cal F}^2_\omega(t) 
         {\cal F}^2_\pi(u)}+ \left[ 8(4 m_*^2-s) p_{1\ominus} p_{4\ominus}
          -4 t (p_{1\ominus}^2+p_{4\ominus}^2) +2 u t \right] 
          \times \nonumber \\
& & \left. \phantom{-g_\omega^2 g_\pi^2 {\cal F}^2_\omega(t)
         {\cal F}^2_\pi(u)} \phantom{fat ghost} \times
          \left\{ G_{\omega N}(t)G_\pi^*(u)+G_{\omega N}^*(t)G_\pi(u) \right\}
          \right) \nonumber \\
& &      -g_\omega^2 g_\pi^2 
          {\cal F}^2_\omega(u) {\cal F}^2_\pi(t) \left(
          \left[ 4 t(2 m_*^2-t) \right] .
          \left\{ G_{\omega T}(u)G_\pi^*(t)+G_{\omega T}^*(u)G_\pi(t) \right\}
          \right. \nonumber \\
& &  \phantom{-g_\omega^2 g_\pi^2 {\cal F}^2_\omega(u) 
          {\cal F}^2_\pi(t)} 
          + \left[ 8(4 m_*^2-s) p_{1\oplus} p_{t\oplus}
          -4 u (p_{1\oplus}^2+p_{3\oplus}^2) +2 u t \right] \times \nonumber \\
& & \left. \phantom{-g_\omega^2 g_\pi^2 {\cal F}^2_\omega(u)
          {\cal F}^2_\pi(t)} \phantom{fat ghost} \times
          \left\{ G_{\omega N}(u)G_\pi^*(t)+G_{\omega N}^*(u)G_\pi(t) \right\}
          \right) \nonumber \\
& & \nonumber \\
& & \nonumber \\
{\cal S}_{2\pi 1\sigma 1\omega} & = & {\cal S}_{2\pi 1\sigma 1\omega}^{d} +
      {\cal S}_{2\pi 1\sigma 1\omega}^{e} + {\cal S}_{2\pi 1\sigma 1\omega}^{i}
      \nonumber \\
& & \nonumber \\
{\cal S}_{2\pi 1\sigma 1\omega}^{d} & = & 0 \nonumber \\
{\cal S}_{2\pi 1\sigma 1\omega}^{e} & = & 0 \nonumber \\
{\cal S}_{2\pi 1\sigma 1\omega}^{i} & = & -g_\sigma g_\omega g_\pi^2
       \left( {\cal F}_\sigma(t){\cal F}_\omega(t) {\cal F}^2_\pi(u)
       \left[ 8 m_* u (p_{1\ominus}+p_{4\ominus}) \right] .
       \left\{ G_\times(t) G_\pi^*(u) +G_\times^*(t) G_\pi(u) \right\}
       \right. \nonumber \\
& & \left. \phantom{-g_\sigma g_\omega g_\pi^2 }
     {\cal F}_\sigma(u){\cal F}_\omega(u) {\cal F}^2_\pi(t)       
       \left[ 8 m_* t (p_{1\oplus}+p_{3\oplus}) \right] .
       \left\{ G_\times(u) G_\pi^*(t) +G_\times^*(u) G_\pi(t) \right\}
       \right) \nonumber 
\eeq

\vskip 1cm

{\bf Acknowledgement}

The authors would like to acknowledge the hospitality of 
GSI, Darmstadt, Germany and Subatech, Nantes, France where part of this
work was performed.

\vskip 1cm

{\bf Figure captions}

\vskip 0.2cm

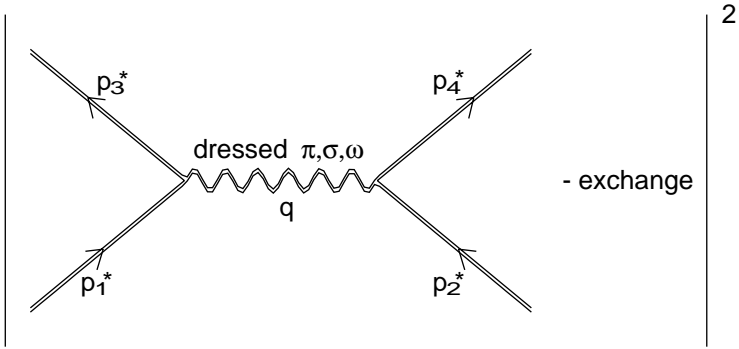
{\bf Fig. 1}: N-N scattering through exchange of one dressed $\pi$, $\sigma$, 
$\omega$ meson

{\bf Fig. 2}: Fitting the experimental total elastic $pp$ cross section:
comparison of the performance of the various parameter sets. Experimental
data points are indicated by the diamonds. The data set A, which is used for
all calculations in this paper, corresponds to the full line.

{\bf Fig. 3}: Fitting the experimental differential elastic cross section:
{\it (a)} comes from a smoothed version of the compilation of existing 
experimental data for five values of the beam energy $T_{\rm lab}$ = 100,
300, 500, 700 and 900 MeV. The dashed part on the left hand side
of the figure corresponds to the range where Coulomb scattering is 
dominant and which was not taken into account for fitting. {\it (b)}
Result of the fit for parameter set A and five values of the beam energy. 
{\it (c)} Differential cross section at saturation density and vanishing 
temperature, for five values of the beam energy.

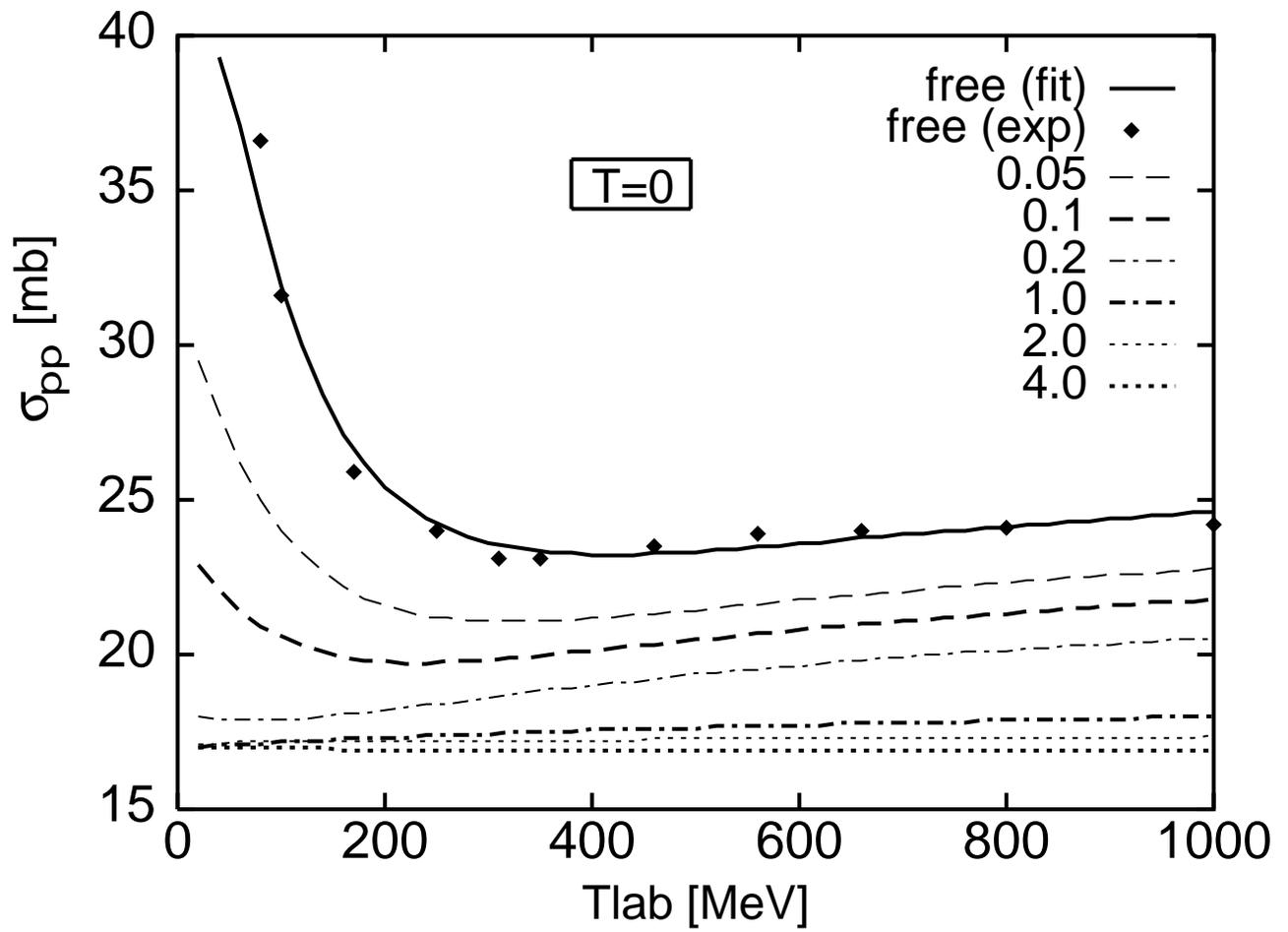
{\bf Fig. 4}: Total elastic $p$-$p$ cross section as a function 
of the beam energy at T=0, several densities. From top to bottom: 
$n/n_{sat}$ = 0, 0.05, 0.1, 0.2, 0.5, 1., 2., 4.

{\bf Fig. 5}: Ratio of the value of the in-medium cross section to its  
value in the vacuum as a function of density $n/n_{sat}$ (log. scale) 
for seven values of the temperature $T$=0, 40, 80, 120, 160, 200 and 240 MeV.
$\sigma_{\rm med}/\sigma_{\rm free}$ is shown on {\it (5a)} for a beam energy
$T_{\rm lab}$ = 100 MeV and on {\it (5b)} for $T_{\rm lab}$ = 300 MeV. 

{\bf Fig. 6}: {\it (a)} Ratio of the value of the in-medium 
cross section to its value in the vacuum as a function of temperature
for five values of the density $n/n_{sat}$=0., 0.1, 1., 2. and 4. and for
two values of the beam energy $T_{\rm lab}$=100 MeV (full line) and
$T_{\rm lab}$=300 MeV (dashed line). {\it (b)}, The effective mass
of the nucleon as a function of temperature is shown for comparison.

{\bf Fig. 7}: Ratio of the value of the in-medium cross section to its 
free value: Contour plot in the density-temperature $log_{10} (n/n_{sat})\ 
{\rm vs.}\ T$ plane, for three values of the energy of the incident particles
$T_{\rm lab}$=100, 300 and 500 MeV.

{\bf Fig. 8}: Differential elastic $p$-$p$ cross section at $T$ = 0 as a 
function of the scattering angle, for a value of the beam energy $T_{\rm lab}$ 
= 800 MeV, and several densities:  $n/n_{sat}$ = 0, 0.05, 0.2, 
0.5, 1., 2., 4.

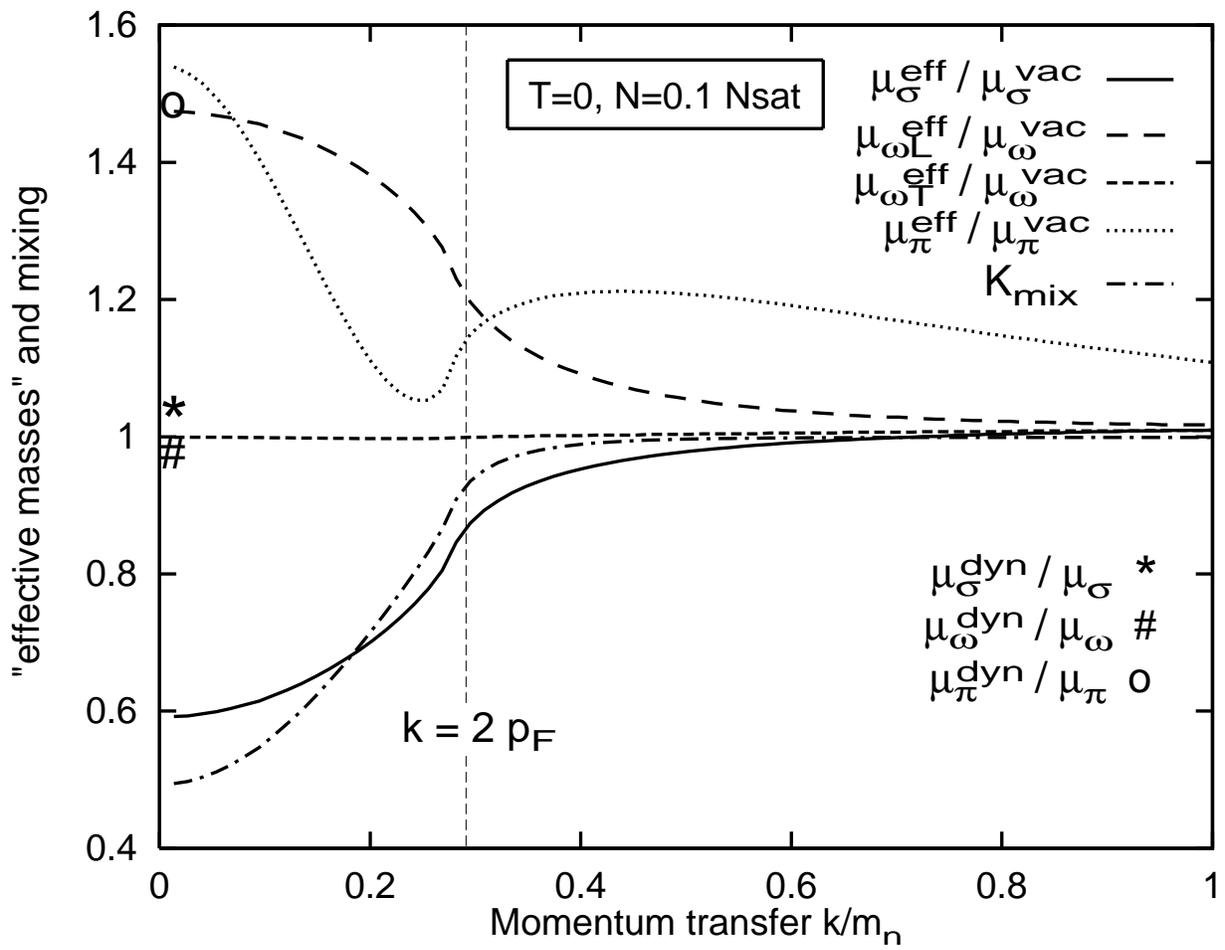
{\bf Fig. 9}: Ratio of the ``effective masses'' (as defined in Eq. 27) 
to their value in the vacuum, and mixing parameter (Eq. 29), as a function 
of momentum transfer at T=0, $n/n_{sat}=0.1$. There is a critical value 
of the momentum transfer $q=2 p_F$. Also shown are the values of the
dynamical masses obtained by solving the dispersion relation at zero 3-momentum
$k=(\omega_i^{\rm pl},\vec 0)$: $\mu_i^{\rm dyn}=\omega_i^{\rm pl}$, $D(k)=0$.

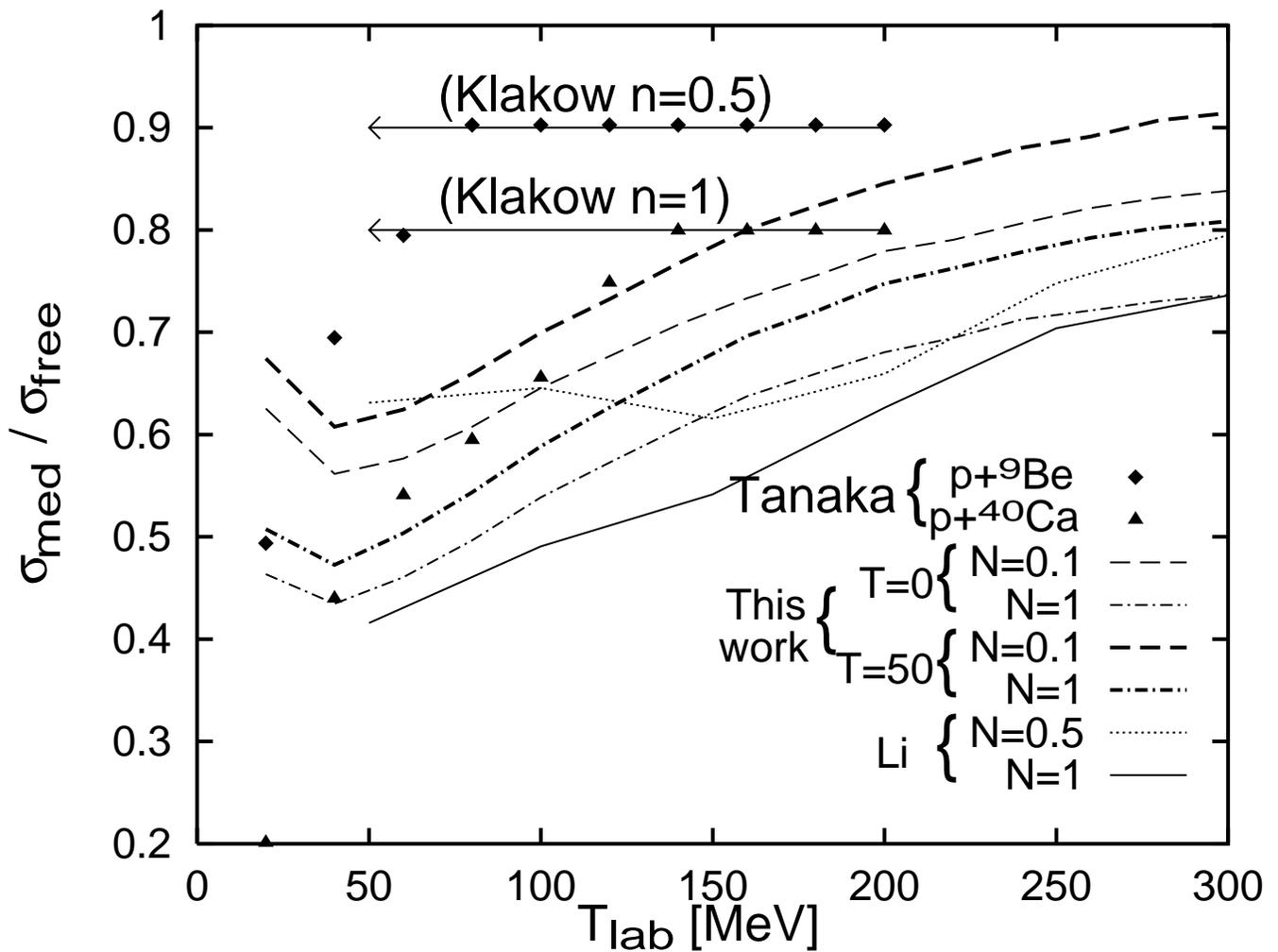
{\bf Fig. 10}: Comparison of experimental constraints on the reduction 
factor on the in-medium cross section from BUU and AMD, Brueckner 
calculations, and our model. The BUU calculation is represented
by the calculation of Klakow and is indicated by two arrows for densities 
$n/n_{sat}$=1. and 0.5. The AMD calculation is represented by the results 
of a simulation by Tanaka {\it et al.} of $p$+${}^9$Be (diamonds) and 
$p$+${}^{40}$Ca (triangles) collisions, thought to probe densities
$n/n_{sat}\ \simeq$  0.1 and 1. respectively. Theoretical results
calculating the medium effects in the Brueckner approach are taken 
from the paper of Li and Machleidt at densities $n/n_{sat}$ =  
0.5 (dotted line) and 1. (full line) and vanishing temperature.
The results of this work are displayed for two values of the temperature
$T$ = 0 (thin lines) and $T$ = 50 MeV (thick lines) and two values of 
the density $n/n_{sat}$ =0.1 (dashed lines) and $n/n_{sat}$ =1. (dash-dotted
line).

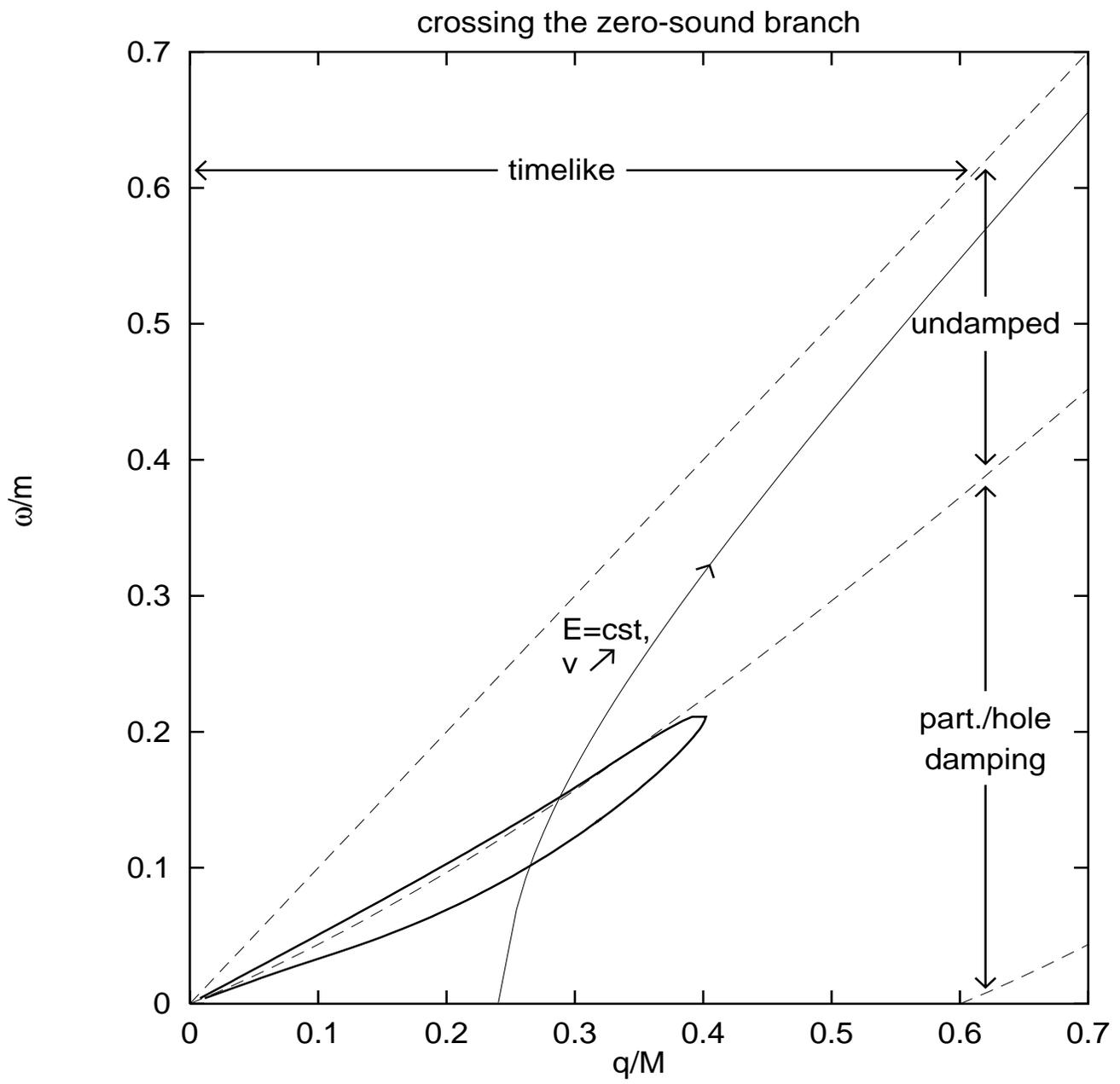
{\bf Fig. 11}: The $\omega$-$q$ plane, particle/hole damping region and 
zero sound branches. The zero-sound branches are calculated for $g_{\sigma}$
= 3.8, $g_{\omega}$ = 9.3 without form factors. For a fixed value of the 
beam energy, scattering angle and quadri-momentum transfer, and by varying 
the relative velocity existing between the rest system of the fluid and the 
center of mass of the collision, the point of coordinates ($\omega$-$q$) 
describes a line which may twice cross the zero sound branches.

\end{quote}

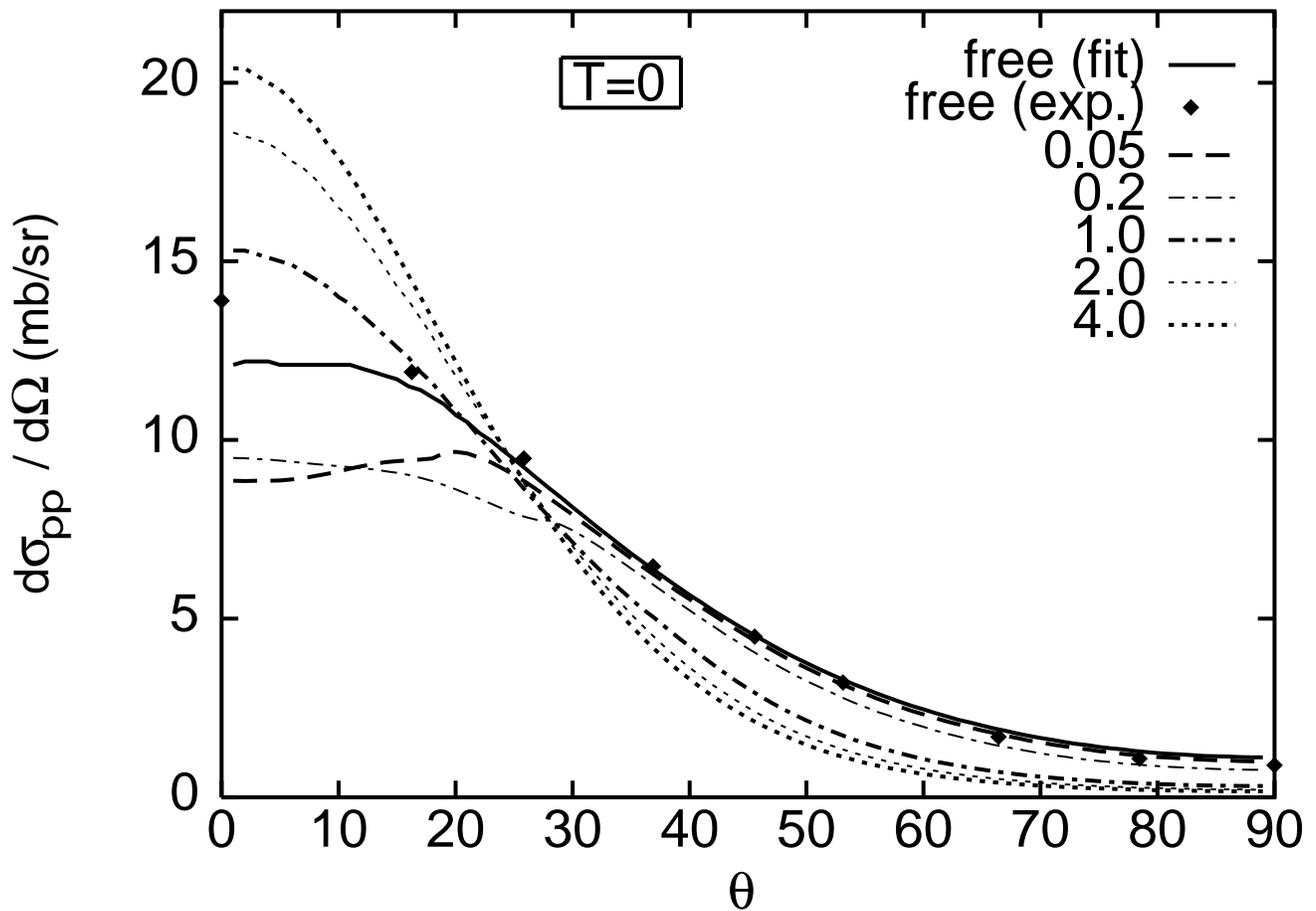
\end{document}